\documentclass[twocolumn,aps,amssymb,floatfix,superscriptaddress,preprintnumbers]{revtex4}
\usepackage[usenames]{color}

\usepackage{graphicx}
\usepackage{bm}
\usepackage{amsmath}
\usepackage{amssymb}
\usepackage{amsfonts}
\usepackage{float}
\usepackage{dsfont}  
\usepackage{slashed}  
\usepackage{booktabs}
\usepackage{multirow}
\usepackage{subfigure}
\usepackage{slashed}
\usepackage[sort&compress]{natbib}

\usepackage{xr-hyper}

\usepackage{hyperref}

\makeatletter
\newcommand*{\addFileDependency}[1]{
  \typeout{(#1)}
  \@addtofilelist{#1}
  \IfFileExists{#1}{}{\typeout{No file #1.}}
}
\makeatother

\newcommand*{\myexternaldocument}[1]{
    \externaldocument{#1}
    \addFileDependency{#1.tex}
    \addFileDependency{#1.aux}
}

\myexternaldocument{Supplementary_material}

\newcommand{\be}{\begin{equation}}  
\newcommand{\ee}{\end{equation}}  
\newcommand{\beq}{\begin{eqnarray}}  
\newcommand{\eeq}{\end{eqnarray}}

\newcommand{\bea}{\begin{eqnarray}}
\newcommand{\eea}{\end{eqnarray}}

\newcommand{\MSb}{{\overline{\rm MS}}}

\begin{document}

\title{Unpolarized and helicity  generalized parton distributions \\[1.5ex] of the proton within lattice QCD}

\author{Constantia Alexandrou}
\affiliation{
Department of Physics,
  University of Cyprus,
  P.O. Box 20537,
  1678 Nicosia,
  Cyprus}
\affiliation{
  Computation-based Science and Technology Research Center,
  The Cyprus Institute,
  20 Kavafi Street,
  Nicosia 2121,
  Cyprus}
\author{Krzysztof Cichy}
\affiliation{Faculty of Physics, Adam Mickiewicz University, Uniwersytetu Pozna\'nskiego 2, 61-614 Pozna\'{n}, Poland}
\author{Martha Constantinou}
\affiliation{Department of Physics,  Temple University,  Philadelphia,  PA 19122 - 1801, USA}
\author{Kyriakos Hadjiyiannakou}
\affiliation{
Department of Physics,
  University of Cyprus,
  P.O. Box 20537,
  1678 Nicosia,
  Cyprus}
\author{Karl Jansen}
\affiliation{NIC, DESY,
  Platanenallee 6,
  D-15738 Zeuthen,
  Germany}
\author{Aurora Scapellato}
\affiliation{Faculty of Physics, Adam Mickiewicz University, Uniwersytetu Pozna\'nskiego 2, 61-614 Pozna\'{n}, Poland}
\author{Fernanda Steffens}
\affiliation{Institut f\"ur Strahlen- und Kernphysik, Rheinische
  Friedrich-Wilhelms-Universit\"at Bonn, Nussallee 14-16, 53115 Bonn
  \vspace*{-0.25cm}
  \centerline{\includegraphics[scale=0.19]{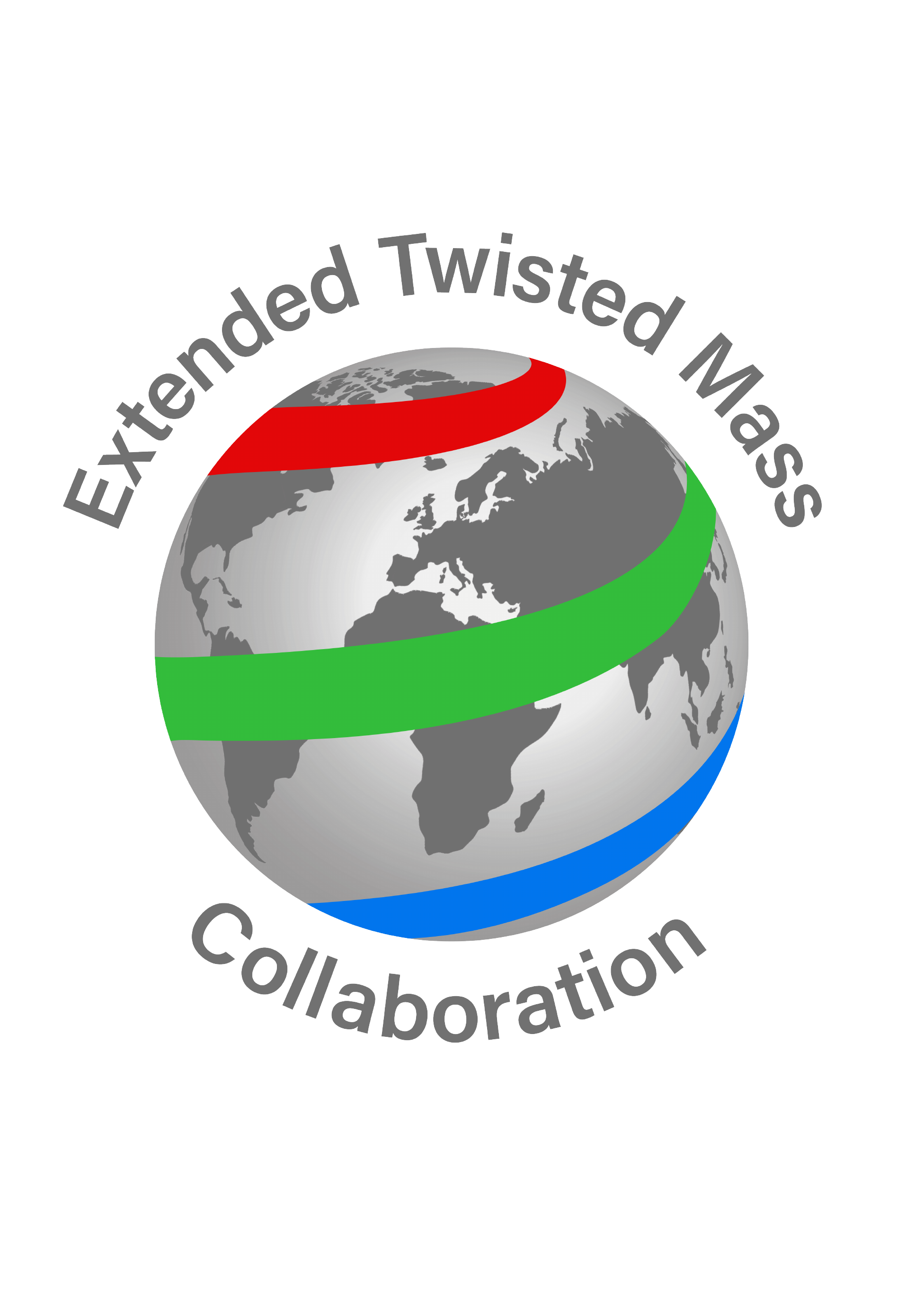}}
   }
  
\begin{abstract}
\vspace*{-.7cm}
\noindent We present the first calculation of the $x$-dependence of the proton generalized parton distributions (GPDs) within lattice QCD. Results are obtained for the isovector unpolarized and helicity GPDs. We compute the appropriate matrix elements of fast-moving protons coupled to non-local operators containing a Wilson line. We present results for proton momenta $0.83,\,1.25,\,1.67$ GeV, and momentum transfer squared $0.69,\,1.38$ GeV$^2$. These combinations include cases with zero and nonzero skewness. The calculation is performed using one ensemble of two degenerate mass light, a strange and a charm quark of maximally twisted mass fermions with a clover term. The lattice results are matched to the light-cone GPDs using one-loop perturbation theory within the framework of large momentum effective theory. The final GPDs are given in the $\MSb$ scheme at a scale of 2 GeV.

\vspace*{0.75cm}

\end{abstract}

\maketitle

\noindent
\textit{Introduction.}
Quantum Chromodynamics (QCD) is the fundamental theory describing the strong interactions among quarks and gluons (partons). The strong force is responsible for binding partons into hadrons, such as the proton, that makes the bulk of the visible matter in the universe. Studying how the properties of protons emerge from the underlying constituents and their interactions has been an important experimental and theoretical endeavor since the mid-20$^{\rm th}$ century. These studies led to the realization that high-energy scattering processes can be factorized into perturbative and non-perturbative parts.
The latter includes information about the parton structure of the proton~\cite{Collins:1989gx}. This resulted in the introduction of a complete set of key quantities, namely the parton distribution functions (PDFs)~\cite{Collins:1989gx}, generalized parton distributions (GPDs)~\cite{Ji:1996ek,Radyushkin:1996nd,Mueller:1998fv}, and transverse momentum dependent distributions (TMDs)~\cite{Collins:2003fm,Ji:2004wu}. These describe the non-perturbative dynamics of the proton, and in general hadrons, in terms of their constituent quarks and gluons~\cite{Collins:1981uw}. 


There are two unpolarized GPDs, $H^q(x,\xi,t)$ and $E^q(x,\xi,t)$, and two helicity GPDs, $\widetilde{H}^q(x,\xi,t)$ and $\widetilde{E}^q(x,\xi,t)$. The superscript $q$ refers to a given quark flavor, and here we study the isovector combination $u-d$. GPDs are functions not only of the longitudinal momentum fraction $x$ ($0\leq x \leq 1$) carried by the partons, but also of the skewness $\xi\equiv-\Delta^+/2P^+$ and the momentum transfer squared, $t\equiv \Delta^2$. $\Delta^+$ and $P^+$ are the plus component of the momentum transfer and the average proton momentum, respectively. Two kinematical regions arise based on the values of $\xi$ and $x$: the so-called Dokshitzer-Gribov-Lipatov-Altarelli-Parisi (DGLAP) region~\cite{Dokshitzer:1977sg,Gribov:1972ri,Lipatov:1974qm,Altarelli:1977zs} defined for $x>|\xi|$, and the Efremov-Radyushkin-Brodsky-Lepage (ERBL)~\cite{Efremov:1979qk,Lepage:1980fj} region for $x<|\xi|$. Physical content can be attributed to each region~\cite{Ji:1998pc} using light-cone coordinates and the light-cone gauge. In the positive- (negative-) $x$ DGLAP region, the GPDs correspond to the amplitude of removing a quark (antiquark) of momentum $k$ from the hadron, and then inserting it back with momentum $k +\Delta$. In the ERBL region, the GPD is the amplitude for removing a quark-antiquark pair with momentum $-\Delta$.

While GPDs are multidimensional objects, they lead to simpler quantities when certain limits are taken, or when integrating over selected variables. For example, the forward limit of the unpolarized case, $\Delta = 0$, gives the quark, $f_1(x)= H^q (x,0,0) $, and antiquark PDFs, $\overline{f}_1(x) = -H^q(-x,0,0)$. Equivalently, in the helicity case one has $g_1(x)=\widetilde{H}^q (x,0,0) $ and $\overline{g}_1(x)=\widetilde{H}^q(-x,0,0)$.  Integrating over $x$ for nonzero $\Delta$, GPDs give the usual FFs. 
Taking integrals of GPDs over $x$ leads to a tower of Mellin moments that also have a physical interpretation. 
such as the total angular momentum of quarks using Ji's sum rule~\cite{Ji:1996ek}.

The connection of GPDs with other quantities demonstrates the information they encode, in both coordinate and momentum spaces. GPDs are accessed through deeply virtual Compton scattering (DVCS) and deeply virtual meson production (DVMP)~\cite{Ji:1996nm}. Despite their importance, it is very difficult to extract them experimentally, even though data are available since the early 2000's. These data are limited, covering a small kinematic region, and are indirectly related to GPDs through the Compton FFs. This poses limitations in their extraction, and the fact that more than one independent measurements are needed to disentangle them~\cite{Diehl:2003ny,Ji:2004gf,Belitsky:2005qn,Kumericki:2016ehc}. 

Nevertheless, the interest in GPDs is renewed due to the advances both on the experimental and the theoretical side, as well as the expertise gained from recent studies of PDFs. It is, thus, of utmost importance to have {\it ab initio} computations of GPDs, that will help map them over different regions of $x$, $\xi$, and $t$. Lattice QCD is the only known formulation that allows a quantitative study of QCD directly using its Lagrangian. Lattice QCD is based on a discretization of Euclidean spacetime and relies on large-scale simulations.

Since parton distributions are light-cone correlation functions \cite{Collins:2011zzd}, it is not straightforward to calculate them using the Euclidean lattice formulation of QCD. The large momentum effective theory (LaMET) proposed by Ji~\cite{Ji:2013dva} provides a promising theoretical framework to extract light-cone quantities using matrix elements computed in lattice QCD. Within LaMET~\cite{Ji:2014gla,Xiong:2013bka}, one can access light-cone quantities via matrix elements of boosted hadrons coupled with non-local spatial operators, which are calculable on the lattice, and yield what is referred to as quasi-distributions. 
The first investigations led to encouraging results on the determination of PDFs~\cite{Lin:2014zya,Alexandrou:2015rja}. Since then, the method has been advanced and attracted a lot of attention, see e.g.\ Refs.~\cite{Chen:2016utp,Alexandrou:2016jqi,Briceno:2017cpo,Constantinou:2017sej,Alexandrou:2017huk,Ji:2017rah,Ji:2017oey,Ishikawa:2017faj,Green:2017xeu,Wang:2017qyg,Stewart:2017tvs,Izubuchi:2018srq,Alexandrou:2018pbm,Alexandrou:2018eet,Chen:2018fwa,Briceno:2018lfj,Spanoudes:2018zya,Alexandrou:2018eet,Liu:2018uuj,Radyushkin:2018nbf,Zhang:2018diq,Li:2018tpe,Alexandrou:2019lfo,Wang:2019tgg,Chen:2019lcm,Izubuchi:2019lyk,Cichy:2019ebf,Wang:2019msf,Son:2019ghf,Green:2020xco,Chai:2020nxw,Braun:2020ymy,Bhattacharya:2020cen,Bhattacharya:2020xlt,Bhattacharya:2020jfj,Chen:2020arf,Chen:2020iqi,Chen:2020ody,Ji:2020brr}, and revitalized other approaches~\cite{Liu:1993cv,Detmold:2005gg,Braun:2007wv,Bali:2017gfr,Bali:2018spj,Detmold:2018kwu,Liang:2019frk}, as well as gave rise to the development and investigation of new ones~\cite{Ma:2014jla,Ma:2014jga,Radyushkin:2016hsy,Chambers:2017dov,Radyushkin:2017cyf,Orginos:2017kos,Ma:2017pxb,Radyushkin:2017lvu,Radyushkin:2018cvn,Zhang:2018ggy,Karpie:2018zaz,Sufian:2019bol,Joo:2019jct,Radyushkin:2019owq,Joo:2019bzr,Balitsky:2019krf,Radyushkin:2019mye,Sufian:2020vzb,Joo:2020spy,Bhat:2020ktg,Can:2020sxc,Alexandrou:2020uyt,Bringewatt:2020ixn} (for recent reviews, see Refs.~\cite{Cichy:2018mum,Ji:2020ect,Constantinou:2020pek}).
Recently, a preliminary study of nucleon GPDs was also presented, demonstrating the applicability of  the quasi-distribution methodology to GPDs~\cite{Alexandrou:2019dax}. The quasi-GPDs approach has also been studied using the scalar diquark spectator model~\cite{Bhattacharya:2018zxi,Bhattacharya:2019cme}. 

\vspace*{0.25cm}
\noindent
\textit{Extracting GPDs using lattice QCD.}
\noindent For the calculation  of GPDs, we define quasi-distributions with boosted proton states and introduce  momentum transfer (denoted $Q$ in Euclidean spacetime) between the initial and final states. 
The  matrix element of interest is given by
\begin{equation}
\hspace*{-0.18cm}
    h_\Gamma(z,P_3,t,\xi) {\equiv} \langle N(P_f)|\bar\psi\left(z\right)\Gamma W(0,z)\psi\left(0\right)|N(P_i)\rangle\,,
\end{equation}
where $|N(P_i)\rangle$ ($|N(P_f)\rangle$) is the initial (final) state labeled by its momentum, and $t=-Q^2$.  For simplicity, we drop the index $q$, since in this work we only consider isovector quantities. 
The boost is in the direction of the Wilson line ($W(0,z)$), $P_3=({P_i}_3+{P_f}_3)/2$. 
Quasi-GPDs depend on the quasi-skewness, defined as $\xi=-\frac{{P_f}_3-{P_i}_3}{{P_f}_3+{P_i}_3}{=}-\frac{Q_3}{2P_3}$ and equal to the light-cone skewness up to power corrections. The Dirac structure $\Gamma$ defines the type of GPD, and we employ $\gamma_0$  and $\gamma_5\gamma_3$ for the unpolarized and helicity GPDs, respectively~\footnote{The operator $\gamma_3$ (unpolarized) is no longer used as it mixes with a twist-3 distribution~\cite{Constantinou:2017sej}}. 

Another aspect of the calculation is the renormalization, as the divergences with respect to the regulator must be removed prior to applying Eq.~(\ref{eq:qGPD}). We adopt~\footnote{For an alternative prescription see Ref.~\cite{Green:2017xeu}} the non-perturbative renormalization scheme of Refs.~\cite{Constantinou:2017sej, Alexandrou:2017huk}, and refined in Ref.~\cite{Alexandrou:2019lfo}. This procedure removes all divergences, including the power-law divergence with respect to the ultraviolet cutoff. The renormalization functions, $Z_\Gamma$, are obtained non-perturbatively by imposing RI-type~\cite{Martinelli:1994ty} renormalization conditions, given in Eq.~(\ref{renorm}). In a nutshell, the final values of $Z_\Gamma$ are obtained at each value of $z$ separately, at a chosen RI scale $(a\mu_0)^2$. For each value of $z$ at a given $\mu_0$, we take the chiral limit using a linear fit in $m_\pi^2$. As described in the supplement, the available matching equations~\cite{Liu:2019urm} require that the quasi-GPDs are in the RI scheme. Therefore, we renormalize the matrix elements using the estimates for $Z_\Gamma$ in the RI scheme at a given scale, $\mu_0$, chosen to be $(a \mu_0)^2\approx1.17$. This scale enters the matching kernel, which converts the quasi-GPDs to light-cone GPDs. The latter are always given in the $\MSb$ scheme at 2 GeV, regardless of the scheme used for quasi-GPDs. Within this work, we explored a few values of the scale within the range $(a\mu_0)^2 \in [1-5]$. We find that the dependence on $(a\mu_0)^2$ is within the reported uncertainties.

The renormalized matrix elements are decomposed into the form factors $\{F_H,\, F_E\}$ and $\{F_{\widetilde{H}},\, F_{\widetilde{E}}\}$, for the unpolarized and helicity case, respectively. The decomposition is based on continuum parametrizations, which in Euclidean space take the form
\begin{eqnarray}
\label{eq:unpol_decomp}
 \langle N(P_f)|{\cal O}_{\gamma_\mu}(z)|N(P_i)\rangle \hspace*{-0.1cm}&=&\hspace*{-0.1cm} \langle\langle\gamma_\mu\rangle\rangle F_H(z,P_3,t,\xi)\nonumber \\ &-&\hspace*{-0.1cm}i\frac{\langle\langle\sigma_{\rho\,\mu}\rangle\rangle\, Q_\rho}{2m} F_E(z,P_3,t,\xi)\,,\hspace*{0.4cm} \\[3ex]
 \label{eq:pol_decomp}
\hspace*{-0.4cm} \langle N(P_f)|{\cal O}_{\gamma_\mu\gamma_5}(z)|N(P_i)\rangle \hspace*{-0.1cm}&=&\hspace*{-0.1cm} \langle\langle\gamma_\mu\gamma_5\rangle\rangle F_{\widetilde{H}}(z,P_3,t,\xi)\nonumber \\ &-&\hspace*{-0.1cm}i\langle\langle\gamma_5\rangle\rangle\frac{Q_\mu}{2m} F_{\widetilde{E}}(z,P_3,t,\xi)\,,\hspace*{0.4cm}
\end{eqnarray}
where $Q\equiv P_f-P_i$, and $m$ is the proton mass.
${\cal O}_\Gamma(z)\equiv\bar\psi\left(z\right)\Gamma W(0,z)\psi\left(0\right)$ and $\langle\langle \Gamma \rangle\rangle \equiv \bar{u}_N(P_f,s')\, \Gamma \,u_N(P_i,s)$ with $u_N$ the proton spinors.

The matrix elements $h_\Gamma(z,P_3,t,\xi)$ depend on $z$, which varies from zero up to the half of the spatial extent $L$ of the lattice. One way to reconstruct the $x$-dependence of the GPDs is via a standard Fourier transform, e.g., we define the quasi $H$-GPD as $H_q$:
\begin{equation}
\label{eq:qGPD}
H_q(x,t,\xi,\mu_0,P_3) =
\int \frac{dz}{4\pi} \, e^{-i x P_3 z} \,F_H(z,P_3,t,\xi,\mu_0)\,.
\end{equation}
This simple Fourier transform suffers from an ill-defined inverse problem~\cite{Karpie:2018zaz}. One alternative reconstruction technique that we adopt here is the Backus-Gilbert (BG) method~\cite{BackusGilbert} that leads to a uniquely reconstructed quasi-distribution from the available set of matrix elements. More  details can be found in the supplement.

The matching formula is available to one-loop level in perturbation theory,  for general skewness~\cite{Liu:2019urm}\footnote{For older work on the matching of quasi-GPDs in the transverse momentum cutoff scheme, see Refs.~\cite{Ji:2015qla,Xiong:2015nua}}. In fact, in the limit of $\xi\rightarrow0$, one recovers the matching equations for quasi-PDFs. Furthermore, the matching kernels of $H$- and $E$-GPDs are the same~\cite{Liu:2019urm}. We provide details on the matching in the supplement.

\vspace*{0.25cm}
\noindent
\textit{Numerical techniques.}
\noindent For this calculation, we employ an ensemble with two light, a strange and a charm quark ($N_f=2+1+1$) using the  twisted mass formulation~\cite{Frezzotti:2000nk,Frezzotti:2003ni} with clover improvement~\cite{Sheikholeslami:1985ij}, generated by the Extended Twisted Mass Collaboration (ETMC)~\cite{Alexandrou:2018egz}. The ensemble has a spatial (temporal) extent of 3 fm (6 fm) ($32^3\times64$), a lattice spacing of 0.093 fm and pion mass of about 260 MeV. For the isovector combination $u-d$, we need to evaluate only the connected diagram (see Fig.~\ref{fig:diagram}).

To increase the signal-to-noise ratio, we use momentum smearing~\cite{Bali:2016lva}, which has been very successful in the calculation of matrix elements of non-local operators with boosted hadrons~\cite{Alexandrou:2016jqi,Alexandrou:2018pbm,Alexandrou:2018egz,Alexandrou:2019lfo}. We find that momentum smearing decreases the gauge noise of the real (imaginary) part by a factor of 4-5 (2-3) (see, e.g., Fig.~\ref{fig:momentum_smearing}). To further suppress statistical uncertainties, we apply stout smearing~\cite{Morningstar:2003gk} to the links of the operator. The effectiveness of the stout smearing in proton matrix elements was demonstrated in Refs.~\cite{Alexandrou:2016ekb,Alexandrou:2020sml}. While the stout smearing changes the matrix elements, it also alters $Z_\Gamma$, and the renormalized matrix elements are independent of the stout smearing.

Ensuring ground-state dominance in $h_\Gamma$ is essential and is controlled by the time separation between the source (initial state) and the sink (final state). This separation, $t_s$, needs to be large in order to suppress excited-states contributions to the matrix elements. We construct a suitable ratio of two- and three-point functions (see Eq.~(\ref{Eq:ratio})), to cancel out unknown overlap factors. Multiple ratios are obtained, for each operator insertion time $t_{ins}=1,\ldots,t_s-a$ (assuming the source time is zero). Ground-state dominance is established when the ratio becomes time independent for values of $t_{\rm ins}$ (plateau region) that are far away enough from the source and the sink (see Eq.~(\ref{Eq:ratio2})). The matrix elements $h_\Gamma(z,P_3,t)$ are extracted from a constant fit within the plateau region. Here, we choose $t_s=1.12$ fm~\cite{Alexandrou:2019lfo}, and use the sequential method at fixed $t_s$ value.

The most common definition of GPDs is in the Breit frame, in which the momentum transfer $Q$ is equally shared between the initial and final states. This has important implications for the computational cost of extracting $h_\Gamma(z,P_3,t,\xi)$ as compared to the usual FFs. For different momentum transfers, both the source and the sink momenta change, requiring separate inversions for each value of $Q$. The statistics used for the results presented in this work is given in Tabs.~\ref{tab:stat}-\ref{tab:stat_PDFs}. We note that, for the largest value of proton momentum, $P_3=1.67$~GeV, the number of measurements required to reach sufficient accuracy is 112\,192. The supplement contain more information on the technical aspects and includes Refs~\cite{Albanese:1987ds,Gusken:1989qx,Alexandrou:1992ti,Gockeler:1998ye,Alexandrou:2015sea,Constantinou:2010gr,Tikhonov:1963,Ulybyshev:2017szp,Ulybyshev:2017ped,Alexandrou:2013joa}.

\vspace*{0.25cm}
\noindent
\textit{Results for the matrix elements $h_\Gamma$.}
\noindent The renormalized matrix elements are decomposed into $F_H,\,F_E,\,F_{\widetilde H}$, and $F_{\widetilde E}$ using Eqs.~(\ref{eq:unpol_decomp})-(\ref{eq:pol_decomp}). To disentangle $F_H$ and $F_E$, we use $h_{\gamma^0}(z,P_3,t,\xi)$ projected with the unpolarized projector, ${\cal P}_0\equiv (1+\gamma^0)/4$ and the polarized projector, ${\cal P}_\kappa \equiv (1+\gamma^0) i \gamma^5 \gamma^\kappa/4$. 
For the helicity matrix element, $h_{\gamma^3\gamma^5}(z,P_3,t,\xi)$, we use the polarized  projector, ${\cal P}_\kappa$, where both $\kappa=3$ and $\kappa\ne3$ are necessary to disentangle $F_{\widetilde H}$ and $F_{\widetilde E}$. We note that for zero skewness, only $\kappa=3$ leads to a non-zero matrix element for the axial vector operator, which is related to $F_{\widetilde H}$. Thus, for $\xi=0$ we cannot access the $\widetilde{E}$-GPD. In fact, the inaccessibility of $\widetilde{E}$-GPD is a general feature due to its vanishing kinematic factor at $\xi=0$, and is not related to the choice of the projector.

For the largest momentum, $P_3=1.67$~GeV, we find similar magnitude contributions from both projectors ${\cal P}_0$ and ${\cal P}_1$. These matrix elements are combined to solve a system of linear equations to extract $F_H$ and $F_E$. Due to its kinematic coefficient, $F_E$ has, in general, larger errors than those for $F_H$. We find that the momentum dependence changes based on the values of $z$, and on the quantity under study. This momentum dependence propagates in a nontrivial way to the final $H$- and $E$-GPDs, as one has to reconstruct the quasi-GPDs in momentum space, and then, apply the appropriate matching formula, which depends on the momentum $P_3$. The matrix element $h_{\gamma^5\gamma^3}$ at zero skewness leads directly to $F_{\widetilde{H}}$, as the kinematic factor of $\widetilde{E}$ is zero. More details and plots can be found in the supplement.

\vspace*{0.25cm}
\noindent
\textit{Results on the GPDs.}
\noindent
The $P_3$-convergence of the GPDs is of particular interest, as the matching kernel is only known to one-loop level. For $H$-GPD and $\widetilde{H}$-GPD at $\xi=0$, we find that the momentum dependence is small and within the reported uncertainties. Convergence is also observed for $E$-GPD for the two highest momenta and the region $x>0$. We note that the statistical errors on $E$-GPD are larger than those of the $H$-GPD, a feature already observed in $F_E$. We refer the Reader to the supplement for more details.

\begin{figure}[h]
\centering
\includegraphics[scale=0.6]{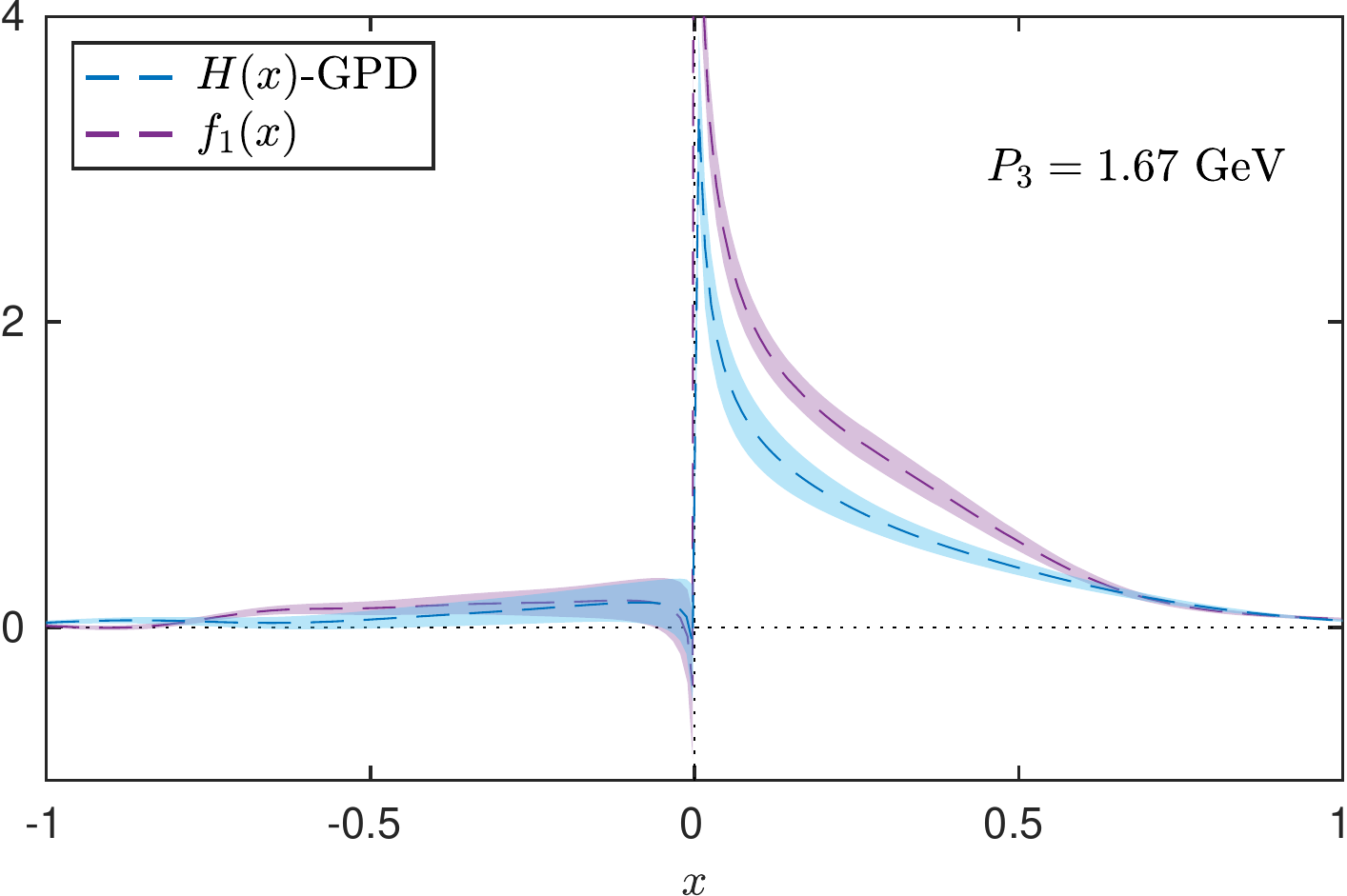}
\vspace*{-0.5cm}
\caption{$H$-GPD (blue band) and unpolarized PDF (violet band) for $P_3=1.67$ GeV and zero skewness.}
\label{fig:H_vs_PDF}
\end{figure}

\begin{figure}[h]
\centering
\includegraphics[scale=0.6]{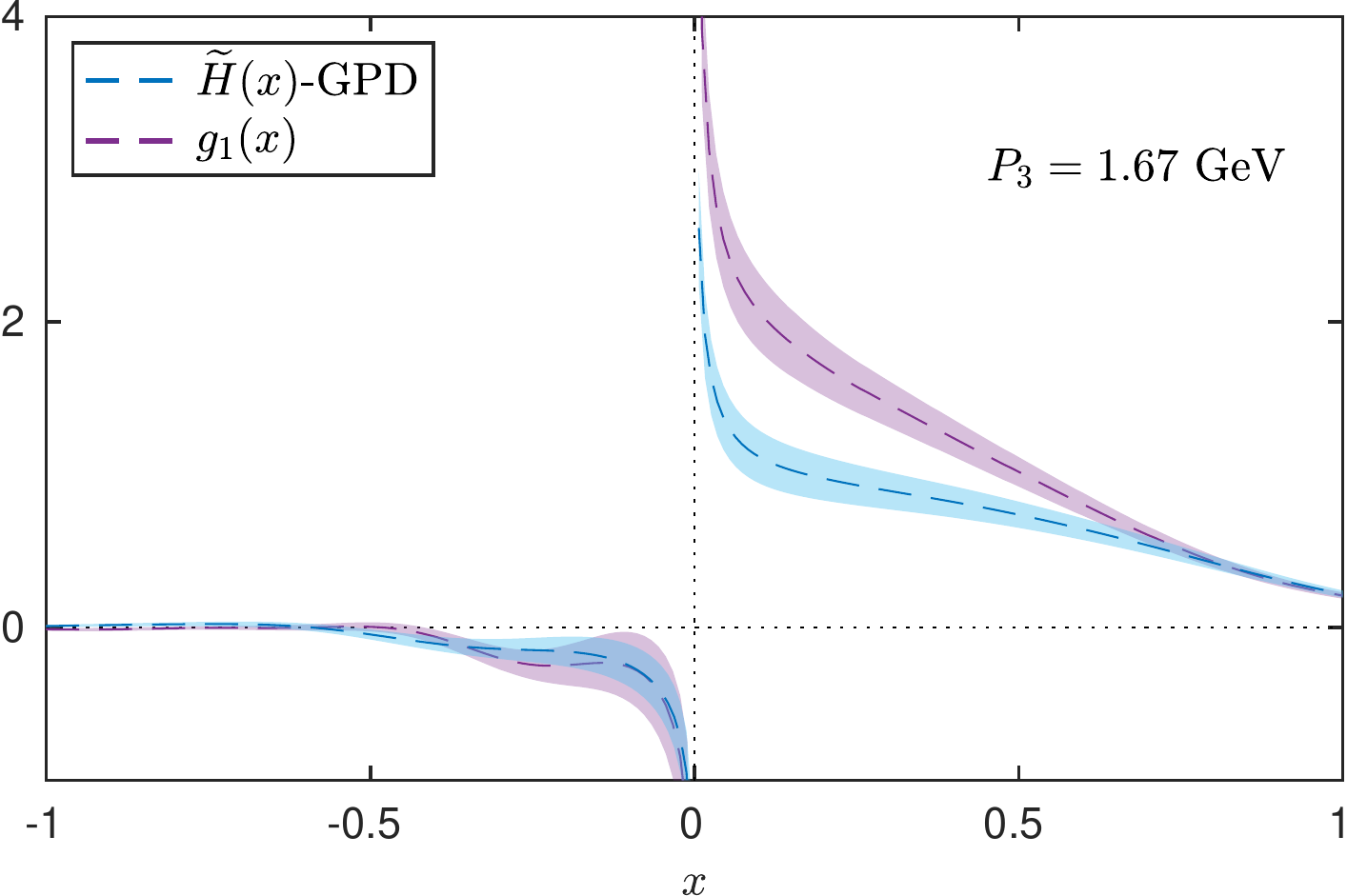}
\vspace*{-0.5cm}
\caption{$\widetilde{H}$-GPD (blue band) and helicity PDF (violet band) for $P_3=1.67$ GeV and zero skewness.}
\label{fig:Htilde_vs_PDF}
\end{figure}

Our final results for $P_3=1.67$ GeV, $t=-0.69$ GeV$^2$, and zero skewness are shown in Fig.~\ref{fig:H_vs_PDF} and Fig.~\ref{fig:Htilde_vs_PDF} for the unpolarized and helicity GPDs, respectively. For each case, we compare the GPDs with the corresponding PDFs, that is $f_1(x)$ for the unpolarized, and $g_1(x)$ for the helicity. We observe that the GPDs are suppressed in magnitude as compared to their respective PDFs for all values of $x\lesssim0.7$. In fact, $\widetilde{H}$-GPD has a steeper slope at small $x$ values. The smaller magnitude of the GPDs is a feature also observed in the standard FFs, which decay with increasing $-t$. For the large-$x$ region, both distributions decay to zero in the same way. The large-$x$ behavior of the unpolarized GPD is in agreement with the power counting analysis of Ref.~\cite{Yuan:2003fs}. For the antiquark region, we find that the GPDs are compatible with the corresponding PDFs. We note that the statistical uncertainties of GPDs are similar to the PDFs, allowing for such qualitative comparison. 

\begin{figure}[h]
\centering
\includegraphics[scale=0.6]{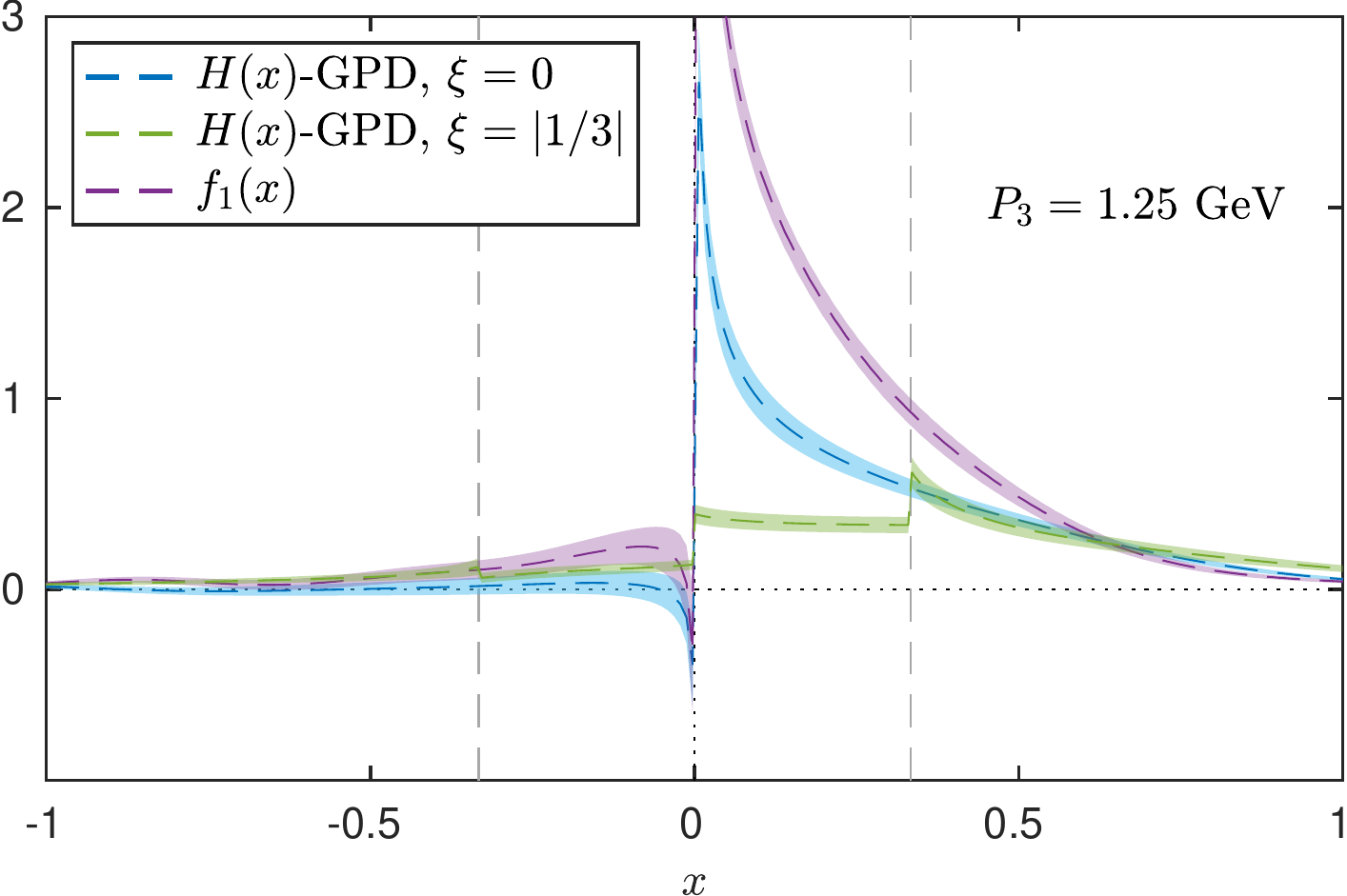}
\vspace*{-0.5cm}
\caption{$H$-GPD for $\xi=0$ (blue band) and $\xi=|1/3|$ (green band), as well as the unpolarized PDF (violet band) for $P_3=1.25$ GeV. The area between the vertical dashed lines is the ERBL region.} 
\label{fig:H_vs_PDF_p2}
\end{figure}

\begin{figure}[h]
\centering
\includegraphics[scale=0.6]{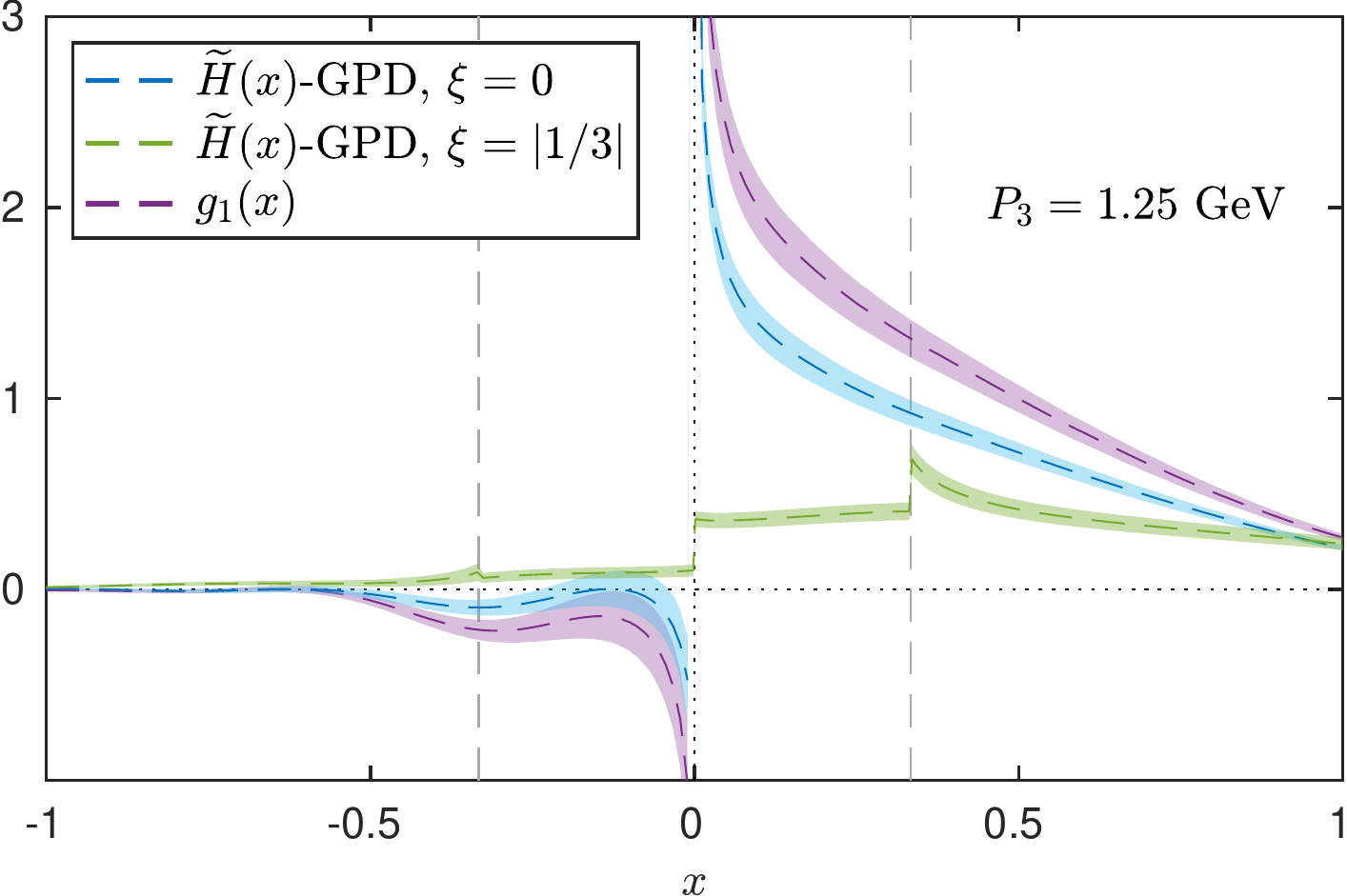}
\vspace*{-0.5cm}
\caption{$\widetilde{H}$-GPD for $\xi=0$ (blue band) and $\xi=|1/3|$  (green band), as well as the helicity PDF (violet band) for $P_3=1.25$ GeV. The area between the vertical dashed lines is the ERBL region.} 
\label{fig:Htilde_vs_PDF_p2}
\end{figure}

The extraction of the GPDs for $\xi \ne 0$ differs from the one for $\xi=0$, as a different matching kernel is required. Also, unlike the $\xi=0$ case, both helicity GPDs contribute to the matrix element, and therefore a decomposition is required. The comparison between the zero and non-zero skewness is shown in Fig.~\ref{fig:H_vs_PDF_p2} and Fig.~\ref{fig:Htilde_vs_PDF_p2}, for $P_3=1.25$ GeV. The main feature of the GPDs at $\xi\ne0$ is that an ERBL region ($|x|< 1/3$ in our case) appears, differentiating it from the DGLAP region ($|x|> 1/3$). The behavior of the GPDs as a function of $t$ for a fixed $x$ is as expected; increasing $-t$ suppresses the GPDs.

\vspace*{0.25cm}
\noindent
\textit{Concluding remarks.}
\noindent
We presented first results on the unpolarized and helicity GPDs for the proton, employing the quasi-distribution approach, which has been very successful for the extraction  of PDFs within lattice QCD. In the case of GPDs, a non-zero momentum is transferred between boosted  initial and final states. The lattice QCD data were renormalized non-perturbatively, and the Backus-Gilbert method was used to extract the $x$-dependence of quasi-GPDs. Applying matching to the latter within the LaMET approach yielded the light-cone GPDs in the $\MSb$ scheme at 2 GeV. 

The momentum dependence of GPDs for $P_3=0.83,\,1.25,\,1.67$ GeV at fixed $t{=}-0.69$ GeV$^2$ (Figs.~\ref{fig:H_E_GPD_mom_depend}, \ref{fig:Htilde_GPD_mom_depend} of the supplementary material) indicates convergence between the largest two momenta. Our final results, given in Figs.~\ref{fig:H_vs_PDF}-\ref{fig:Htilde_vs_PDF} at zero skewness and Figs.~\ref{fig:H_vs_PDF_p2}-\ref{fig:Htilde_vs_PDF_p2} at nonzero skewness, are reassuring, as with increasing $-t$, the magnitude of GPDs is suppressed. With our calculation, we demonstrate that extracting GPDs with controlled statistical uncertainties is feasible. Their accuracy permits qualitative comparison with their corresponding PDFs. 

In the near future, we will investigate systematic uncertainties, as studied for PDFs~\cite{Alexandrou:2019lfo}. The pion mass dependence will also be studied using an ensemble with quark masses fixed to their physical values. In a follow-up calculation, we will also explore the transversity GPD, for which there are two additional form factors, leading to a more evolved decomposition. This makes the disentanglement of the transversity GPDs more challenging.

The current work demonstrates the feasibility of the quasi-distributions approach for GPDs using computational resources that are within reach. However, there is still a long way, until statistical and systematic uncertainties become under control. Extracting GPDs within the \textit{first principles} formulation of lattice QCD can potentially be combined with future experimental data within the global fits framework. This direction is very timely, as GPDs are at the heart of planned experiments at JLab~\cite{Biselli:2018nuv} and the Electron-Ion-Collider (EIC)~\cite{Accardi:2012qut}. Therefore, GPDs are the objects to drive the efforts of the nuclear and hadronic physics communities for the next decades. 

\begin{acknowledgements}
We would like to thank all members of ETMC for their constant and pleasant collaboration. K.C.\ and A.S.\ are supported by National Science Centre (Poland) grant SONATA BIS no.\ 2016/22/E/ST2/00013. M.C. acknowledges financial support by the U.S. Department of Energy Early Career Award under Grant No.\ DE-SC0020405. K.H. is supported by the Cyprus Research and Innovation Foundation under grant POST-DOC/0718/0100. F.S.\ was funded by DFG project number 392578569. Partial support is provided by the European Joint Doctorate program STIMULATE of the European Union’s Horizon 2020 research and innovation programme under grant agreement No. 765048. Computations for this work were carried out in part on facilities of the USQCD Collaboration, which are funded by the Office of Science of the U.S. Department of Energy. This research was supported in part by PLGrid Infrastructure (Prometheus supercomputer at AGH Cyfronet in Cracow).
Computations were also partially performed at the Poznan Supercomputing and Networking Center (Eagle supercomputer), the Interdisciplinary Centre for Mathematical and Computational Modelling of the Warsaw University (Okeanos supercomputer) and at the Academic Computer Centre in Gda\'nsk (Tryton supercomputer). The gauge configurations have been generated by the Extended Twisted Mass Collaboration on the KNL (A2) Partition of Marconi at CINECA, through the Prace project Pra13\_3304 "SIMPHYS". 
\end{acknowledgements}

\newpage
\setcounter{equation}{0}
\setcounter{figure}{0}
\setcounter{table}{0}

\parskip=6pt

\begin{widetext}
\appendix

\setcounter{equation}{0}
\setcounter{figure}{0}
\setcounter{table}{0}

\parskip=6pt

\makeatletter
\renewcommand{\theequation}{S\arabic{equation}}
\renewcommand{\thefigure}{S\arabic{figure}}
\renewcommand{\thetable}{S\Roman{table}}

\section{SUPPLEMENTARY MATERIAL}

\subsection{Lattice Methods}

The work is based on the calculation of proton matrix elements of the nonlocal operator containing a Wilson line in the $z$ direction, $W(0,z)$, that is
\begin{equation}
\label{eq:ME}
h_\Gamma(z,P_3,t)\equiv \langle N(P_f)|\bar\psi\left(z\right)\Gamma W(0,z)\psi\left(0\right)|N(P_i)\rangle\,.
\end{equation}
An important requirement of the quasi-distribution approach is that the hadron is boosted with a momentum in the same direction as the Wilson line, therefore $\mathbf{P}=(0,0,P_3)$. GPDs are multidimensional objects and require momentum transfer between the initial and final states. In Euclidean space, this is defined as $Q^2$, which is related to its Minkowski counterpart as  $t \equiv - Q^2$. An important parameter of GPDs is the skewness, which is proportional to the momentum transfer in the direction of the boost. In the quasi-distribution method, the relevant quantity is the quasi-skewness defined as $\xi\equiv-\frac{Q_3}{2P_3}$.

\begin{figure}[ht]
\centerline{\includegraphics[scale=0.9]{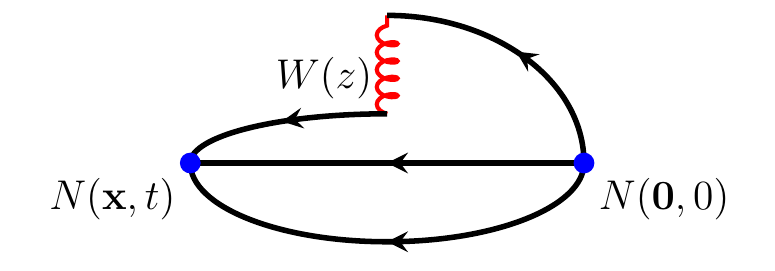}}
\vspace*{-0.15cm}
\begin{minipage}{15cm}
\caption{Connected diagram for the evaluation of the three-point function. The initial and final states for the proton are indicated by $N(\mathbf{0},0)$ and $N(\mathbf{x},t)$, respectively.}
\label{fig:diagram}
\end{minipage}
\end{figure}

We calculate the isovector flavor combination, which requires calculation of only the connected diagram shown in Fig.~\ref{fig:diagram}.
To obtain the ground state of the matrix element, one must calculate two-point and three-point correlation functions,
\be
C^{\rm 2pt}(\mathbf{P},t,0) ={{\cal P}_0}_{\alpha\beta}\sum_\mathbf{x}e^{-i\mathbf{P}\cdot \mathbf{x}}\langle 0\vert N_\alpha(\mathbf{x},t) N_\beta(\mathbf{0},0)\vert 0 \rangle\,, 
\ee
\be
C_{\Gamma}^{\rm 3pt}(\mathbf{P_f},\mathbf{P_i};t,\tau,0) =  {{\cal P}_\kappa}_{\alpha\beta}\,\sum_{\mathbf{x},\mathbf{y}}\,e^{i(\mathbf{P_f}-\mathbf{P_i})\cdot \mathbf{y}}\,e^{-i\mathbf{P_f}\cdot \mathbf{x}} 
 \langle 0\vert N_{\alpha}(\mathbf{x},t) \mathcal{O}_{\gamma^j\,\gamma^5}(\mathbf{y},\tau;z)N_{\beta}(\mathbf{0},0)\vert 0\rangle\,,
\ee
where $N_{\alpha}(x){=}\epsilon ^{abc}u^a _\alpha(x)\left( d^{b^{T}}(x)\mathcal{C}\gamma_5u^c(x)\right)$ is the interpolating field for the proton, $\tau$ is the current insertion time. Without loss of generality, we take the source to be at $(\mathbf{0},0)$. ${\cal P}_0$ is the parity plus projector ${\cal P}_0{=}\frac{1{+}\gamma_4}{2}$, and ${\cal P}_\kappa$ is either ${\cal P}_0$ or ${\cal P}_\kappa=\frac{1}{4}(1+\gamma^0) i \gamma^5 \gamma^j$ if $\kappa$ is spatial. For nonzero momentum transfer, one must form an optimized ratio in order to cancel the time dependence in the exponentials and the overlaps between the interpolating field and the nucleon states,
\begin{equation}
R_{\Gamma}({\cal P}_{\kappa},\mathbf{P_f},\mathbf{P_i};t,\tau) = \frac{C_{\Gamma}^{3pt}({\cal P}_\kappa,\mathbf{P_f},\mathbf{P_i};t,\tau\
)}{C^{2pt}({\cal P}_0,\mathbf{P_f};t)} \times 
 \sqrt{\frac{C^{2pt}({\cal P}_0,\mathbf{P_i};t-\tau) C^{2pt}({\cal P}_0,\mathbf{P_f};\tau) C^{2pt}({\cal P}_0,\mathbf{P_f};t)}{C^{2pt}({\cal P}_0,\mathbf{P_f};t-\tau) C^{2pt}({\cal P}_0,\mathbf{P_i};\tau) C^{2pt}({\cal P}_0,\mathbf{P_i};t)}}\,.
\label{Eq:ratio}
\end{equation}
In the limit $(t_s-\tau) \gg a$ and $\tau \gg a$, the ratio of Eq.(\ref{Eq:ratio}) becomes time-independent and the ground state matrix element is extracted from a constant fit in the plateau region, that is
\begin{equation}
\label{Eq:ratio2}
R_{\Gamma}({\cal P}_\kappa;\mathbf{P_f},\mathbf{P_i};t;\tau)\xrightarrow[\tau\gg a]{t-\tau\gg a}\Pi_{\Gamma}({\cal P}\kappa;\mathbf{P_f},\mathbf{P_i})\,.
\end{equation}
The ground state contribution, $\Pi_{\Gamma}$, is decomposed as shown in Eqs.~(2) - (3) of the main text. The expressions for the unpolarized and helicity cases for the non-vanishing projectors can be written as
\begin{eqnarray}      
{\Pi_{\gamma^0}({\cal P}_0;{P_f},{P_i})} &=&
 C \Bigg[    F_H(Q^2) \left(\frac{ E_f  E_i }{2
    m^2}+\frac{ E_f+ E_i }{4 m}+\frac{{P_f}_\rho \, {P_i}_\rho}{4
    m^2}+\frac{1}{4}\right)  \nonumber \\[2ex]  
  && \hspace*{.4cm} + F_E(Q^2)
    \left({P_f}_\rho\, {P_i}_\rho \left(\frac{ E_f+ E_i }{8
    m^3}+\frac{1}{4 m^2}\right)-\frac{ (E_f- E_i)^2}{8 m^2}+\frac{ E_f+ E_i }{8
    m}+\frac{1}{4}\right)
        \Bigg] \,, \\[7ex]  
{\Pi_{\gamma^0}({\cal P}_j;{P_f},{P_i})} &=&
   i\, \epsilon_{j\,0\,\rho\,\tau} C \Bigg[
   F_H(Q^2)\frac{ {P_f}_\rho\, {P_i}_\tau}{4 m^2} 
+    F_E(Q^2)
    \left(\frac{( E_f+ E_i ) {P_f}_\rho \,{P_i}_\tau}{8
    m^3}+\frac{{P_f}_\rho\,
    {P_i}_\tau - {P_f}_\tau\, {P_i}_\rho}{8 m^2}\right)\Bigg]\,, \\[7ex]  
{\Pi_{\gamma^3\gamma^5}({\cal P}_j;{P_f},{P_i})} &=&
 i \,C\,\Bigg[ F_{\widetilde{H}}(Q^2)  \left(\delta_{j 3}
    \left(\frac{E_f+E_i}{4 m}-\frac{{P_f}_\rho\,\,
    {P_i}_\rho}{4 m^2}+\frac{1}{4}\right)+\frac{{P_i}_3 \,\,{P_f}_j+ {P_f}_3\,\,{P_i}_j}{4
    m^2}\right)   \,\,\,\,\,\,\,\,\, \nonumber \\[2ex]
  && \hspace*{.5cm} -  F_{\widetilde{E}}(Q^2)\frac{ 
    ({P_f}_3-{P_i}_3) (-E_f{P_i}_j+E_i {P_f}_j+m(
    {P_f}_j-{P_i}_j))}{8 m^3}\Bigg]  \,,
\end{eqnarray}
where $C=\frac{2 m^2}{\sqrt{E_f E_i (E_f+m) (E_i+m)}}$, with $E_{i,f} =\sqrt{m^2 + \mathbf{P}_{i,f}^{\,2}}$ . The index $j$ runs only over the spatial components, while a sum over all 4 components is implied for $\rho$ and $\tau$.

To improve the overlap with the proton ground state, we construct the proton interpolating field using momentum-smeared quark fields~\cite{Bali:2016lva}, on APE smeared gauge links~\cite{Albanese:1987ds}. The momentum smearing technique was proven to be crucial to suppress statistical uncertainties for matrix elements with boosted hadrons, and in particular for nonlocal operators~\cite{Alexandrou:2016jqi}. In this work, we can reach $P_3=1.67$ GeV at a reasonable computational cost. The momentum smearing function $\mathcal{S}$ on a quark field, $\psi$, reads
\begin{equation}
\mathcal{S}\psi(x)=\frac{1}{1+6\alpha_G}\left(\psi(x)+\alpha_G\sum_{j=1}^{3}U_j(x)e^{i\overline{\xi}\mathbf{P}\cdot\mathbf{j}}\psi(x+\hat{\mathbf{j}})\right),   
\label{eq:mom_smearing_function}
\end{equation}
where $\alpha_G$ is the parameter of the Gaussian smearing~\cite{Gusken:1989qx,Alexandrou:1992ti}, $U_j$ is the gauge link in the spatial $j$-direction. $\mathbf{P}=(P_1,P_2,P_3)$ is the momentum of the proton (either at the source, or at the sink) and $\overline{\xi}$ is a free parameter that can be tuned so that a maximal overlap with the proton boosted state is achieved. For $\overline{\xi}=0$, Eq.~(\ref{eq:mom_smearing_function}) reduces to the Gaussian smearing function.
In our implementation, we keep $\overline{\xi} \mathbf{P}$ parallel to the proton momentum at the source and at the sink. Such a constraint requires separate quark propagators for every momentum transfer, because the gauge links are modified every time by a different complex phase. However, this strategy avoids potential problems due to rotational symmetry breaking. It also has the benefit that every correlator entering the ratio of Eq.~(\ref{Eq:ratio}) is optimized separately.

\begin{figure}[h!]
\includegraphics[scale=0.54]{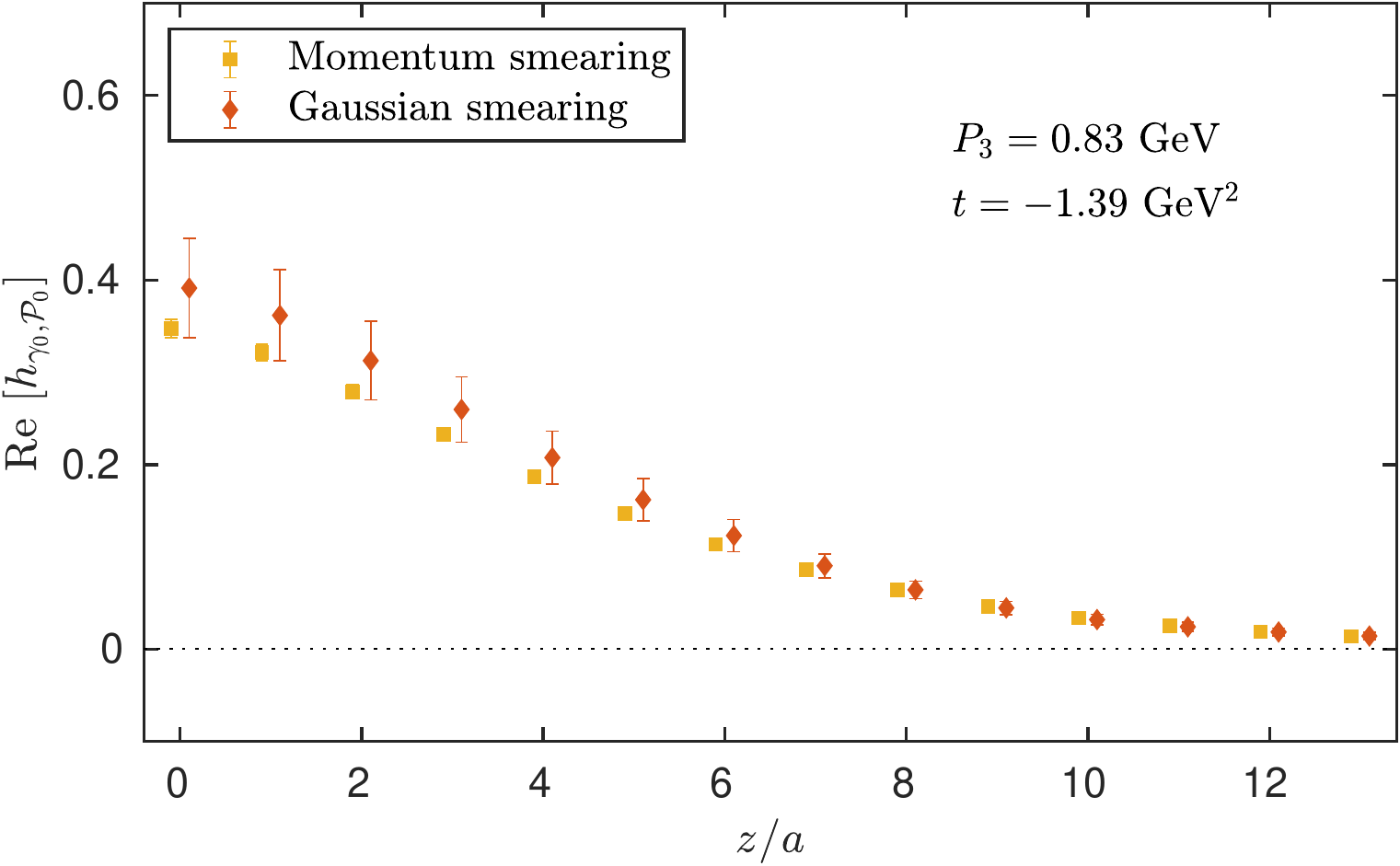}\\[3ex]
\includegraphics[scale=0.54]{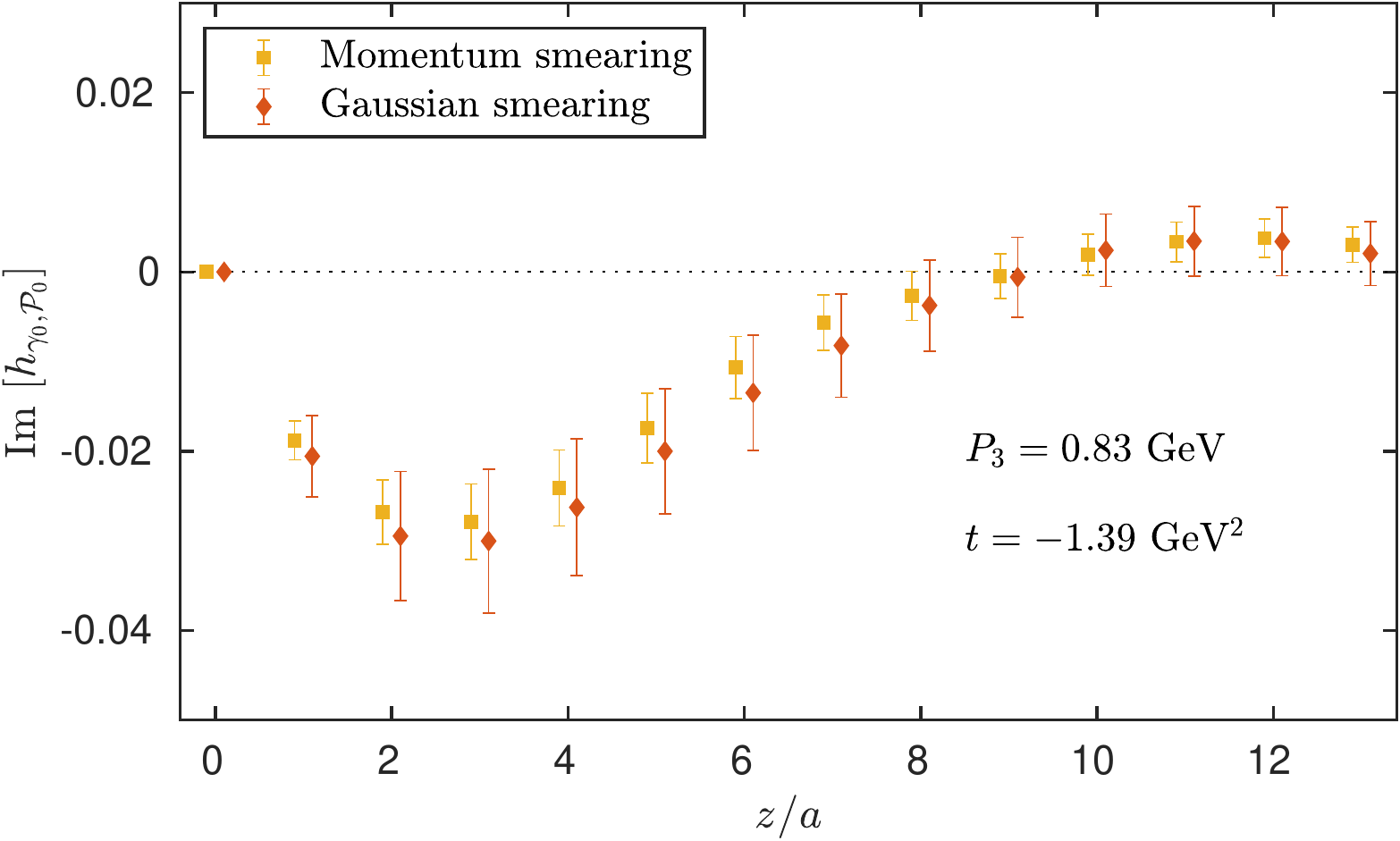}
\begin{minipage}{15cm}
\caption{Bare matrix elements obtained for the Dirac insertion $\gamma_0$, using the unpolarized parity projector, $\Gamma_0$. The data with momentum smearing are shown with yellow squares, while the data using the plain Gaussian smearing are shown with red diamonds. The data correspond to $P_3=0.83$ GeV and $t=-1.39$ GeV$^2$.}
\label{fig:momentum_smearing}
\end{minipage}
\end{figure}

As an example, we show in Fig.~\ref{fig:momentum_smearing} the bare matrix elements of the vector operator with and without momentum smearing. For this comparison, we use the unpolarized parity projector, momentum boost $P_3=0.83$ GeV, and momentum transfer $t=-1.39$ GeV$^2$. The individual components of the momentum transfer are given by the vector $\vec{Q} \frac{L}{2\pi}=(0,-2,2)$. The number of measurements is 1616 for $t_s=8a$, and the data using the momentum smearing correspond to $\overline{\xi}=0.6$ after optimization. As can be seen, the use of momentum smearing significantly decreases the statistical uncertainties for both the real and imaginary parts of the matrix elements. In particular, the statistical accuracy increases by a factor of 4-5 in the real part, and 2-3 in the imaginary part, depending on the value of $z$.

In Table~\ref{tab:stat}, we summarize the statistics for each value of the nucleon momentum $P_3$, the momentum transfer squared $t$, and skewness $\xi$.
In Table~\ref{tab:stat_PDFs}, we give the numbers of measurements for the corresponding PDFs, to which the computed GPDs can be compared. 

\begin{table}[h!]
\begin{center}
\renewcommand{\arraystretch}{1.4}
\begin{tabular}{cccc|cc}
\hline
$P_3$ [GeV] & $\quad \vec{Q}\times \frac{L}{2\pi}\quad$ & $-t$ [GeV$^2$] & $\xi$ & $N_{\rm confs}$ & $N_{\rm meas}$\\
\hline
0.83 &(0,2,0)  &0.69  &0      & 519  & 4152\\
1.25 &(0,2,0)  &0.69  &0      & 1315  & 42080\\
1.67 &(0,2,0)  &0.69  &0      & 1753  & 112192\\
1.25 &(0,2,2)  &1.39  & 1/3   & 417  & 40032\\
1.25 &(0,2,-2) &1.39  & -1/3  & 417  & 40032 \\
\hline
\end{tabular}
\begin{minipage}{15cm}
\caption{Statistics for both unpolarized and helicity GPDs, at each momentum boost, momentum transfer and skewness.}
\label{tab:stat}
\end{minipage}
\end{center}
\end{table}

\begin{table}[h!]
\begin{center}
\renewcommand{\arraystretch}{1.4}
\begin{tabular}{c|c|cc}
\hline
$P_3$ [GeV] & $\Gamma$ & $N_{\rm confs}$ & $N_{\rm meas}$\\
\hline
0.83  & $\gamma_0$ & 115 & 920\\
& $\gamma_5\gamma_3$ & 194 & 1560\\
\hline 
1.25  & $\gamma_0$, $\gamma_5\gamma_3$& 731  & 11696\\
\hline 
1.67  & $\gamma_0$, $\gamma_5\gamma_3$ & 1644  & 105216\\
\hline
\end{tabular}
\begin{minipage}{15cm}
\caption{Statistics for the unpolarized  and helicity PDFs at the three nucleon momenta. The Dirac structures used in three-point functions calculations are $\Gamma=\gamma_0$ and $\Gamma=\gamma_5\gamma_3$ for the unpolarized and helicity distributions, respectively.}
\label{tab:stat_PDFs}
\end{minipage}
\end{center}
\end{table}

\subsection{Renormalization} 

We employ an RI-type renormalization prescription, using the momentum source method~\cite{Gockeler:1998ye,Alexandrou:2015sea} that offers high statistical accuracy. The appropriate conditions for the renormalization functions of the nonlocal operator, $Z_{\Gamma}$, and the quark field, $Z_q$, are
\begin{eqnarray}
\label{renorm}
Z_q^{-1}\, Z_{\Gamma}(z)&& \hspace*{-0.5cm}  {\rm Tr} \left[{\cal V}_{\Gamma}(p,z) \, \slashed{p} \right] \Bigr|_{p^2{=} \mu_0^2} =
{\rm Tr} \left[{\cal V}_{\Gamma}^{{\rm Born}}(p,z)\, \slashed{p} \right] \Bigr|_{p^2{=} \mu_0^2}
\, , \\[3ex]
&& \hspace*{-0.5cm} Z_q =\frac{1}{12} {\rm Tr} \left[(S(p))^{-1}\, S^{\rm Born}(p)\right] \Bigr|_{p^2= \mu_0^2}\,.
\end{eqnarray}
Note that Eq.~(\ref{renorm}) is applied at each value of $z$ separately. ${\cal V}(p,z)$ ($S(p)$) is the amputated vertex function of the operator (fermion propagator) and $S^{{\rm Born}}(p)$ is the tree-level of the propagator. 

The above prescription is mass-independent, and therefore $Z_{\Gamma}$ do not depend on the quark mass. However, there might be residual cut-off effects of the form $a m_q$. To eliminate any systematic related to such effects, we extract $Z_{\Gamma}$ using multiple degenerate-quark ensembles ($N_f=4$) at the same lattice spacing and action as the $N_f=2+1+1$ ensemble used for the production of the nucleon matrix elements. We then take the chiral limit  of the $Z_{\Gamma}$. The exact procedure is described in detail in Ref.~\cite{Alexandrou:2019lfo}. We use five $N_f=4$ ensembles as given in Tab.~\ref{Table:Z_ensembles}, which  have been produced for the calculation of the renormalization functions at the same $\beta$ value as the $N_f=2+1+1$ ensemble used for the extraction of the matrix elements of Eq.~(\ref{eq:ME}). 
\begin{table}[ht]
\begin{center}
\renewcommand{\arraystretch}{1.5}
\renewcommand{\tabcolsep}{5.5pt}
\begin{tabular}{ccc}
\hline\hline 
$\beta=1.726$, & $c_{\rm SW} = 1.74$, & $a=0.093$~fm \\
\hline\hline\\[-3ex]
{$24^3\times 48$}  & {$\,\,a\mu = 0.0060$}  & $\,\,m_\pi = 357.84$~MeV     \\
\hline
{$24^3\times 48$}  & $\,\,a\mu = 0.0080$     & $\,\,m_\pi = 408.11$~MeV     \\
\hline
{$24^3\times 48$}  & $\,\,a\mu = 0.0100$    & $\,\,m_\pi = 453.48$~MeV    \\
\hline
{$24^3\times 48$}  & $\,\,a\mu = 0.0115$    & $\,\,m_\pi = 488.41$~MeV    \\
\hline
{$24^3\times 48$}  & $\,\,a\mu = 0.0130$    & $\,\,m_\pi = 518.02$~MeV    \\
\hline\hline
\end{tabular}
\vspace*{-0.25cm}
\begin{minipage}{15cm}
\caption{\small{Parameters of the $N_f=4$ ensembles used for the calculation of the renormalization functions}.}
\label{Table:Z_ensembles}
\end{minipage}
\end{center}
\end{table} 

The RI renormalization scale $ \mu_0$, defined in Eq.~(\ref{renorm}), is chosen appropriately to have suppressed discretization effects, as explained in Ref.~\cite{Alexandrou:2015sea}. We employ several values that have the same spatial components, that is $p=(p_0,p_1,p_1,p_1)$, so that the ratio $\frac{p^4}{(p^2)^2}$ is less than 0.35, as suggested in Ref.~\cite{Constantinou:2010gr}. In this work, we use different values of $ \mu_0$ ($(a\, \mu_0)^2 \in [1,5]$) to check the dependence of the matching formalism on $ \mu_0$. 
For each $ \mu_0$ value, we apply a chiral extrapolation using the fit
\begin{equation}
\label{eq:Zchiral_fit}
Z^{\rm RI}_{\Gamma}(z, \mu_0,m_\pi) = {Z}^{\rm RI}_{{\Gamma},0}(z, \mu_0) + m_\pi^2 \,{Z}^{\rm RI}_{{\Gamma},1}(z, \mu_0) \,,
\end{equation}
to extract the mass-independent ${Z}^{\rm RI}_{{\Gamma},0}(z, \mu_0)$.

As mentioned in the main text, the matching kernel of Ref.~\cite{Liu:2019urm} requires that the quasi-GPDs are renormalized in the RI scheme. For consistency, we use the same
for $\xi=0$ and use the renormalization functions defined on a single ${\rm RI}$ renormalization scale, $(a  \mu_0)^2\approx1.17$. This scale also enters the matching equations. We find negligible dependence when varying $ \mu_0$.

\subsection{Reconstruction of $x$-dependence}
\label{sec:reconstruction}
The renormalized matrix elements $F_G$, where $G=H,E$, are related to the  quasi-distributions $G_q$ by a Fourier transform: 
\begin{equation}
\label{eq:X2F}
F_G(z,P_3,t,\xi,\mu_0) = \int_{-\infty}^\infty dx \, e^{i x P_3 z} \, G_q(x,t,\xi,\mu_0,P_3)\,.
\end{equation}
Inverting this expression relates the quasi-GPDs to the matrix elements.
However, the inverse equation involves a Fourier transform over a continuum of lengths of the Wilson line, up to infinity, while the lattice provides only a discrete set of determinations of $F_G$, for integer values of $z/a$ up to roughly half of the lattice extent in the boost direction, $L/2a$.
Thus, the inversion of Eq.~(\ref{eq:X2F}) poses a mathematically ill-defined problem, as argued and discussed in detail in Ref.~\cite{Karpie:2018zaz}.
The inverse problem originates from incomplete information, i.e.\ attempting to reconstruct a continuous distribution from a finite number of input data points.
As such, its solution necessarily requires making additional assumptions that provide the missing information.
These assumptions should be as mild as possible and preferably model-independent -- else, the reconstructed distribution may be biased.

One of the approaches proposed in Ref.~\cite{Karpie:2018zaz} is to use the Backus-Gilbert (BG) method \cite{BackusGilbert}.
The model-independent criterion used in the BG procedure, to choose from among the infinitely many possible solutions to the inverse problem, is that the variance of the solution  with respect to the statistical variation of the input data should be minimal.
The reconstruction proceeds separately for each value of the momentum fraction $x$.
In practice, we separate the exponential of the Fourier transform into its cosine and sine parts, related to the real and imaginary parts of the matrix elements, respectively.
We define a vector $\textbf{a}_K(x)$, where $K$ denotes either the cosine or sine kernel, of dimension $d$ equal to the number of available input matrix elements, i.e.\ $d=z_{\rm max}/a+1$, where $z_{\rm max}/a$ is the maximum length of the Wilson line (in lattice units) used to determine the quasi-distribution.
The BG procedure consists in finding the vectors $\textbf{a}_K(x)$ for both kernels according to the variance minimization criterion.
The vector $\textbf{a}_K(x)$ is an approximate inverse of the cosine/sine kernel function $K(x)$, that is:
\begin{equation}
\Delta(x-x')=\sum_{z/a=0}^{d-1} a_K(x)_{z/a} K(x')_{z/a}\,,
\end{equation}
and $K(x')_{z/a}=\cos(x'P_3z)$ or $K(x')_{z/a}=\sin(x'P_3z)$ are elements of a $d$-dimensional vector of discrete kernel values corresponding to available integer values of $z/a$.
The function $\Delta(x-x')$ is, thus, an approximation to the Dirac delta function $\delta(x-x')$, with the quality of this approximation depending, in practice, on the achievable dimension $d$ at given simulation parameters.

The vectors $\textbf{a}_K(x)$ are found from optimization conditions resulting from the BG criterion.
We refer to Ref.~\cite{Karpie:2018zaz} for their explicit form and here, we just summarize the final result.
We define a $d\times d$-dimensional matrix $\textbf{M}_K(x)$, with matrix elements
\begin{equation}
M_K(x)_{z/a,z'/a}=\int_0^{x_c} dx'\,(x-x')^2 K(x')_{z/a} \, K(x')_{z'/a}+\rho\,\delta_{z/a,z'/a}\,,
\end{equation}
where $x_c$ is the maximum value of $x$ for which the quasi-distribution is taken to be non-zero (i.e.\ its reconstruction proceeds for $x\in[0,x_c]$) and the parameter $\rho$ regularizes the matrix $\textbf{M}_K$.
This regularization, proposed by Tikhonov \cite{Tikhonov:1963}, was put up as one possible way of making $\textbf{M}_K$ invertible \cite{Ulybyshev:2017szp,Ulybyshev:2017ped,Karpie:2018zaz}).
The value of $\rho$ determines the resolution of the method and should be taken as rather small, in order to avoid a bias.
We use $\rho=10^{-3}$, which leads to reasonable resolution and is large enough to avoid oscillations in the final distributions related to the presence of small eigenvalues of $\textbf{M}_K$.
Additionally, we define a $d$-dimensional vector $\textbf{u}_K$, with elements
\begin{equation}
u_{K;z/a}=\int_0^{x_c} dx'\,K(x')_{z/a}.    
\end{equation}
The above mentioned optimization conditions lead to:
\begin{equation}
\textbf{a}_K(x)=\frac{\textbf{M}_K^{-1}(x)\,\textbf{u}_K}{\textbf{u}_K^T\,\textbf{M}_K^{-1}(x)\,\textbf{u}_K}
\end{equation}
and the final BG-reconstructed quasi-distributions are given by
\begin{equation}
G_q(x,t,\xi,\mu_0,P_3)=\frac{1}{2} \sum_{z/a} \left( a_{\rm cos}(x)_{z/a} \, {\rm Re}\, F_G(z,P_3,t,\xi) +  a_{\rm sin}(x)_{z/a} \, {\rm Im}\, F_G(z,P_3,t,\xi) \right)\,.
\end{equation}

\subsection{Matching Procedure}
\label{sec:matching}

Contact between the physical distributions and the quasi-GPDs is established through a perturbative matching procedure. The factorization formula for the Dirac structure $\Gamma$ takes the form
\begin{eqnarray}
\label{eq:matching}
G_q(x,t,\xi,\mu_0,(\mu_0)_3,P_3) &=& \int_{-1}^1 \frac{dy}{|y|}\, C_G \left(\frac{x}{y},\frac{\xi}{y},\frac{\mu}{y P_3},\frac{(\mu_0)_3}{y P_3},r\right) G(y,t,\xi,\mu)+\mathcal{O}\left(\frac{m^2}{P_3^2},\frac{t}{P_3^2},\frac{\Lambda_{\rm QCD}^2}{x^2P_3^2}\right),
\end{eqnarray} 
where $C_G$ is the \textit{matching kernel}, known to one-loop level in perturbation theory, and the involved renormalization scales are: $\mu_0$ -- RI renormalization scale, its $z$-component $(\mu_0)_3$ (with $r=\mu_0^2/(\mu_0)_3^2$), and $\mu$ -- final $\MSb$ scale. This formula establishes that quasi-distributions are equal to light-cone distributions up to power-suppressed corrections (nucleon mass ($m$) corrections and higher-twist corrections). The matching coefficient for the GPDs, was first derived for flavor non-singlet unpolarized and helicity quasi-GPDs in Ref.\ \cite{Ji:2015qla} and for transversity quasi-GPDs in Ref.~\cite{Xiong:2015nua}, using the transverse momentum cutoff scheme. Recently, a matching formula was also derived for all Dirac structures \cite{Liu:2019urm} relating quasi-GPDs renormalized in a variant of the RI/MOM scheme to $\MSb$ light-cone PDFs. In these calculations, it was shown that the matching for GPDs at zero skewness is the same as for PDFs. It was also demonstrated that, to one-loop level, the $H$-type and $E$-type GPDs have the same matching formula.
The matching kernel for a given Dirac structure $\Gamma$ and parton momentum $p_3$ reads%
\begin{align}
\label{e:bare_matching}
C_G\left(\Gamma;x,\xi,\frac{p_3}{\mu},\frac{p_3}{(\mu_0)_3},r\right) = \delta(x-1) &+ \frac{\alpha_s C_F}{2\pi}\left\{
\begin{array}{lc}
G_1(\Gamma;x,\xi)_+				& x<-\xi\\
G_2(\Gamma;x,\xi,p_3/\mu)_+		& |x|<\xi\\
G_3(\Gamma;x,\xi,p_3/\mu)_+		& \xi<x<1\\
-G_1(\Gamma;x,\xi)_+				& x>1
\end{array}\right.\nonumber\\
&-\frac{\alpha_s C_F}{2\pi}\left|\frac{p_3}{(\mu_0)_3}\right|f_{\slashed{p}}\left(\Gamma;\frac{p_3}{(\mu_0)_3}(x-1)+1,r\right)_+,
\end{align}
The functions $G_1, G_2,G_3$ for the matching of bare quasi-GPDs can be found in Ref.~\cite{Liu:2019urm}, while the one-loop RI counterterm $f_{\slashed{p}}$ for the variant that we employ (RI-$\slashed{p}$) is given in Ref.~\cite{Liu:2018uuj}. The plus prescription is defined as
\begin{equation}
    f(x)_+	= f(x) + \delta(x-1)\int dy f(y)
\end{equation}
and it combines the so-called "real" (vertex) and "virtual" (self-energy) corrections.

\vspace*{0.5cm}
\subsection{Results}

In this section, we provide more details for the extracted matrix elements and the final GPDs.

In Fig.~\ref{fig:ME_unpol}, we show the bare matrix elements for the vector operator, using the projectors ${\cal P}_0$ ($h_{\gamma_0,{\cal P}_0}$) and ${\cal P}_1$ with $\kappa=1$ ($h_{\gamma_0,{\cal P}_1}$). Note that for the polarized projector, only $\kappa=1$ contributes, as the momentum transfer has zero component in that direction. Focusing on the largest value of the  momentum $P_3=1.67$~GeV, one observes that both $h_{\gamma_0,{\cal P}_0}$ and $h_{\gamma_0,{\cal P}_1}$ give similar contributions. The decomposition of the renormalized matrix elements leads to $F_H$ and $F_E$, shown in Fig.~\ref{fig:H_E_unpol}. It is interesting to observe that the statistical errors for $F_E$ are, in general, larger than those for $F_H$. This effect has its origin in the kinematic coefficients of $F_E$ in the decomposition of the matrix element.
We find that the momentum dependence changes based on the values of $z$, and on the quantity under study. This momentum dependence propagates in a nontrivial way to the final $H$- and $E$-GPDs, as one has to reconstruct the quasi-GPDs in momentum space, and then, apply the appropriate matching formula, which depends on the momentum $P_3$.

\begin{figure}[h!]
\centering
\includegraphics[scale=0.54]{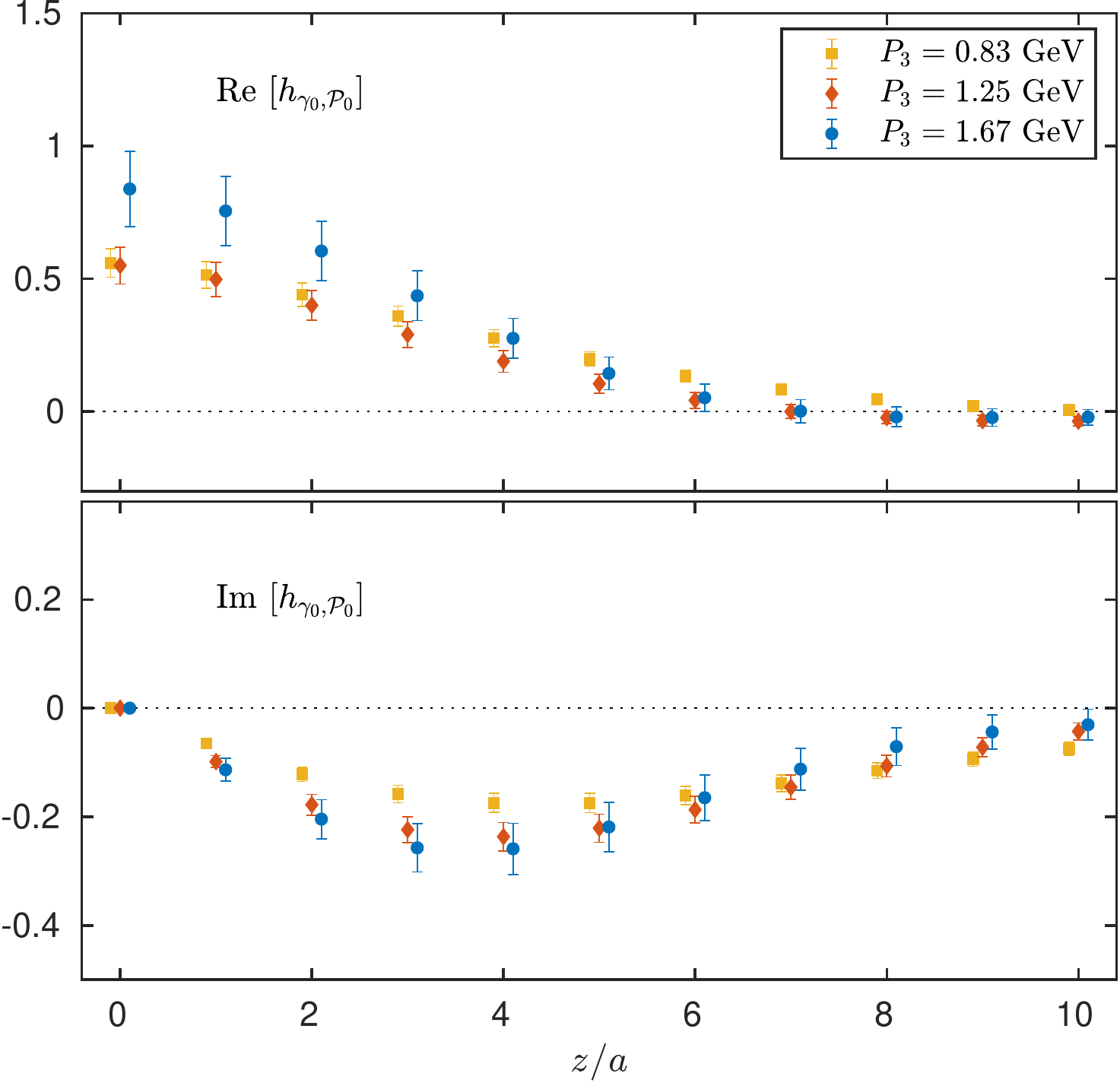}\hspace{0.2cm}
\includegraphics[scale=0.54]{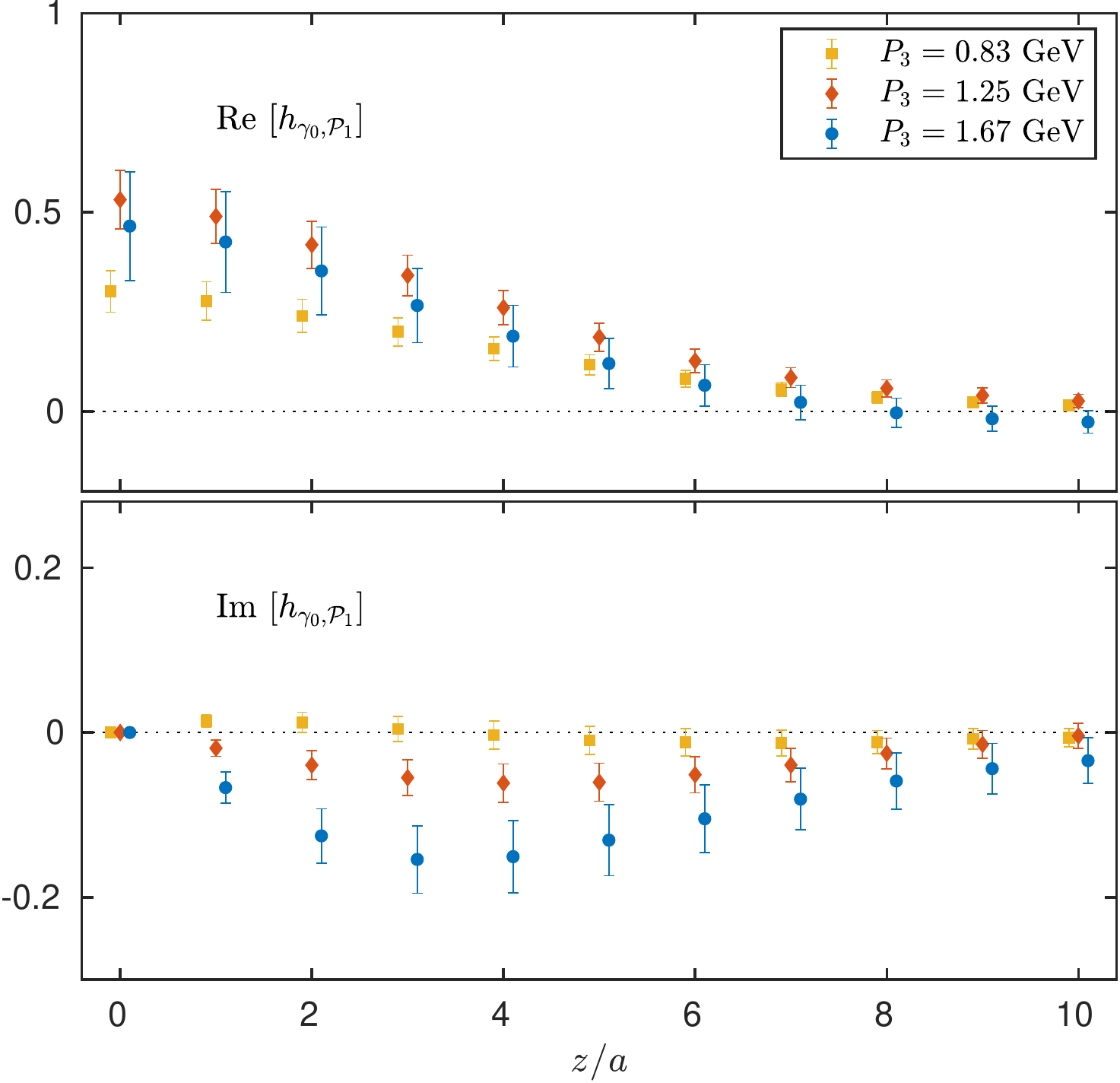}
\vspace*{-0.35cm}
\hspace*{0.35cm}
\begin{minipage}{15cm}
\caption{The bare matrix elements $h_{\gamma_0}$ using parity projector ${\cal P}_0$ (left panels) and  ${\cal P}_1$ (right panels) for zero skewness and $t=-0.69$ GeV$^2$. The top (bottom) plots correspond to the real (imaginary) part of the matrix elements. Momenta $P_3=0.83,\,1.25,\,1.67$ GeV are shown with orange squares, red diamonds, blue circles, respectively.}
\vspace*{0.5cm}
\label{fig:ME_unpol}
\end{minipage}
\end{figure}

\begin{figure}[h!]
\centering
\includegraphics[scale=0.54]{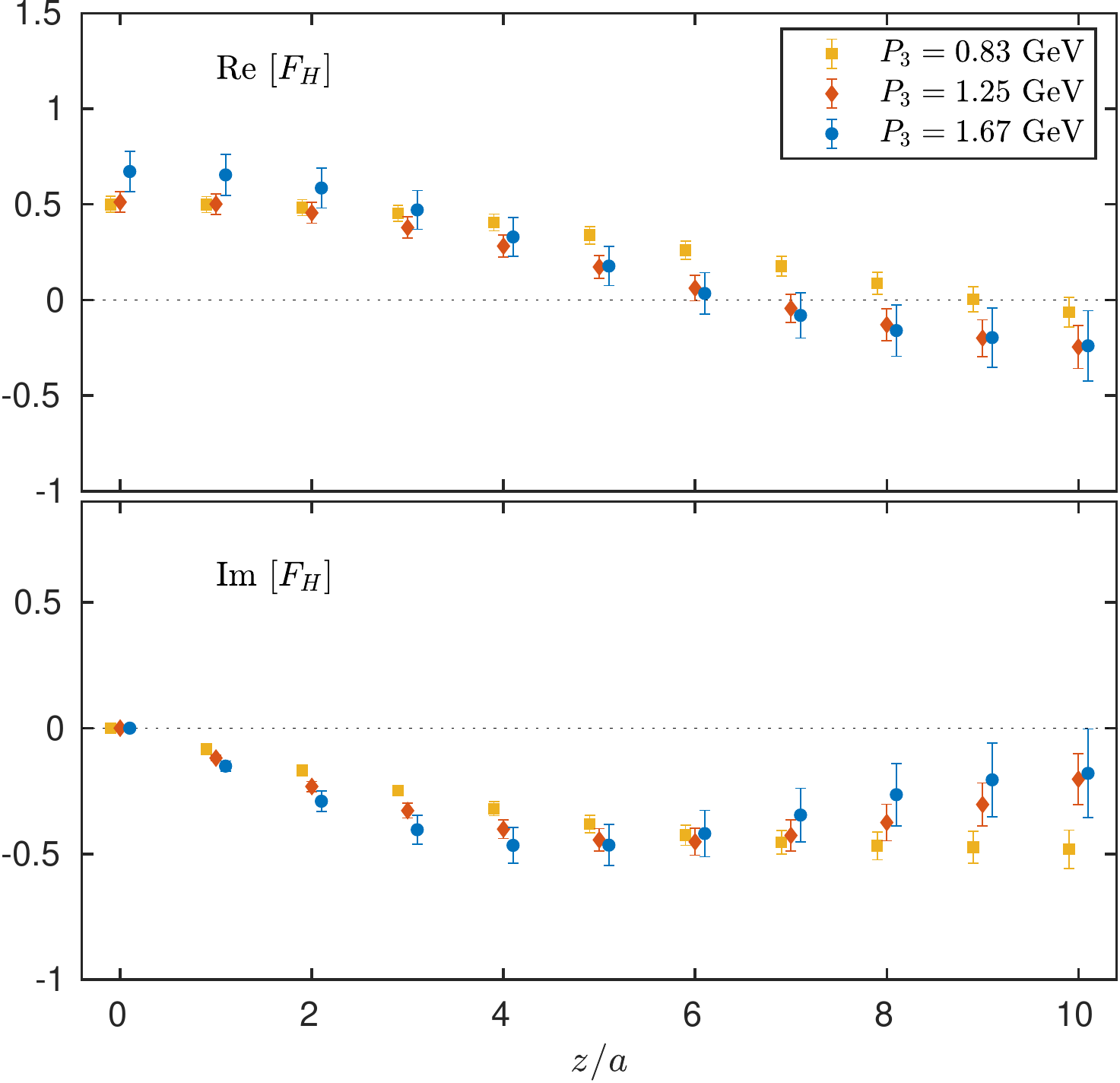}\hspace{0.2cm}
\includegraphics[scale=0.54]{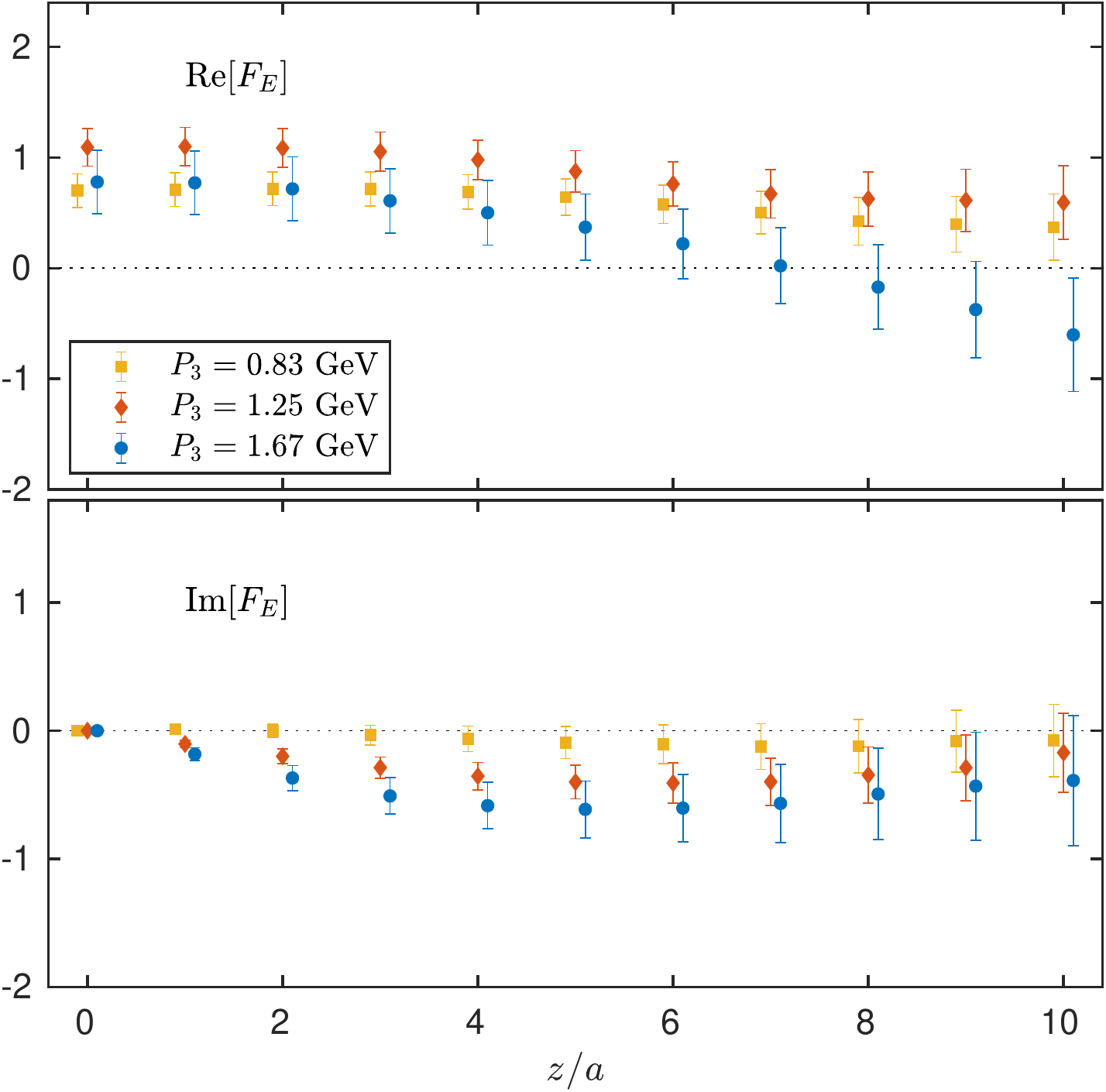}
\vspace*{-0.35cm}
\hspace*{0.35cm}
\begin{minipage}{15cm}
\caption{The renormalized quantities $F_H$ (left panels) and  $F_E$ (right panels) for $t=-0.69$ GeV$^2$ and $\xi=0$. The top (bottom) plots correspond to the real (imaginary) part of the matrix elements. Momenta $P_3=0.83,\,1.25,\,1.67$ GeV are shown with orange squares, red diamonds, blue circles, respectively.}
\vspace*{0.5cm}
\label{fig:H_E_unpol}
\end{minipage}
\end{figure}

In Fig.~\ref{fig:H_E_unpol} we show the decomposed quantities $F_H$ and $F_E$, which have been disentangled using the renormalized matrix elements $h_{\gamma^0,{\cal P}_0}$ and $h_{\gamma^0,{\cal P}_1}$. We find negligible momentum dependence in ${\rm Re}[F_H]$ for $0<z<5$, while it has a steeper  and compatible slope for the two largest momenta in the region $5\leq z \le 9$. ${\rm Re}[F_H]$ flattens out for all three momenta for $z\ge 10$, which are consistent. 
Similarly, ${\rm Im}[F_H]$ is compatible for the largest two momenta for all values of $z$, while the lowest momentum evinces a clearly slower decay to zero.
This might indicate onset of convergence above $P_3=1.25$ GeV. 
Note, however, that the matching depends on $P_3$ and thus, more conclusions about convergence can be drawn after applying this procedure.
For ${\rm Re}[F_E]$, the errors are significantly larger than for ${\rm Re}[F_H]$, as remarked above, and we observe somewhat slower decay at $P_3=1.25$ GeV as compared to $P_3=1.67$ GeV.
This may indicate slower convergence in the $E$-GPD, but may also be a statistical effect, since ${\rm Im}[F_E]$ is, again, compatible for the largest two boosts. 
Similarly to ${\rm Re}[F_H]$, ${\rm Re}[F_E]$ also approaches zero for $z\ge 6$. Finally, we find that ${\rm Im}[F_E]$ has very small contribution for the lowest momentum, while it is enhanced in the intermediate $z$ region for the largest two boosts and comparable in magnitude to ${\rm Im}[F_H]$.

The matrix element $h_{\gamma^3\gamma^5}$ is shown in the left panel of Fig.~\ref{fig:ME_pol} for $t=-0.69$ GeV$^2$ and $\xi=0$. The corresponding $F_{\widetilde{H}}$ is shown in the right panel. We note that for zero skewness, the kinematic factor of $\widetilde{E}$ is zero, and we only extract $\widetilde{H}$ from the lattice QCD data.
We observe that both for the real and the imaginary part of $F_{\widetilde{H}}$, there are significant differences between the largest two momenta.
Thus, we postpone conclusions about convergence to the discussion of the final GPDs.

\begin{figure}[h!]
\vspace*{0.5cm}
\centering
\includegraphics[scale=0.54]{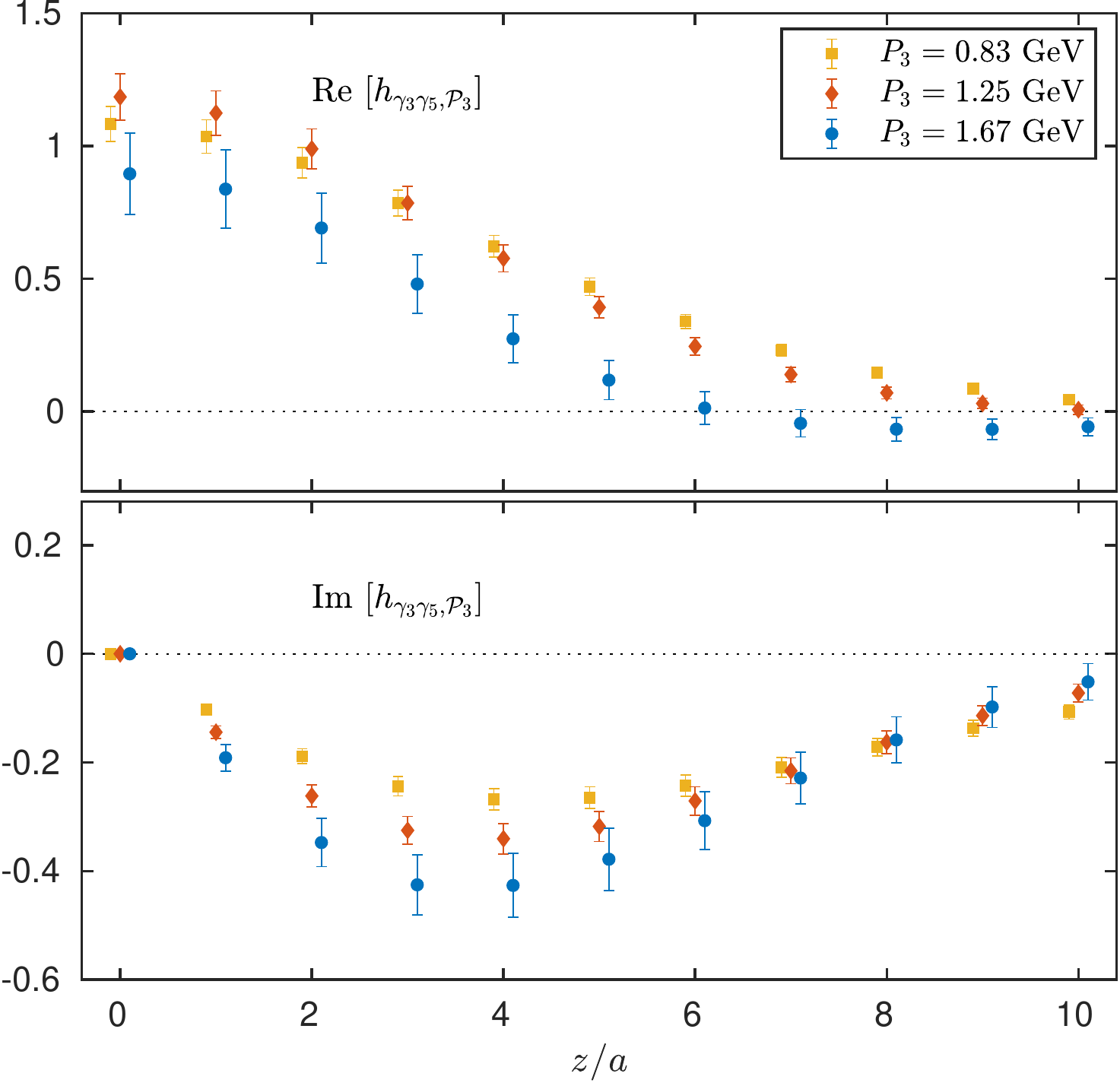}\hspace{0.2cm}
\includegraphics[scale=0.54]{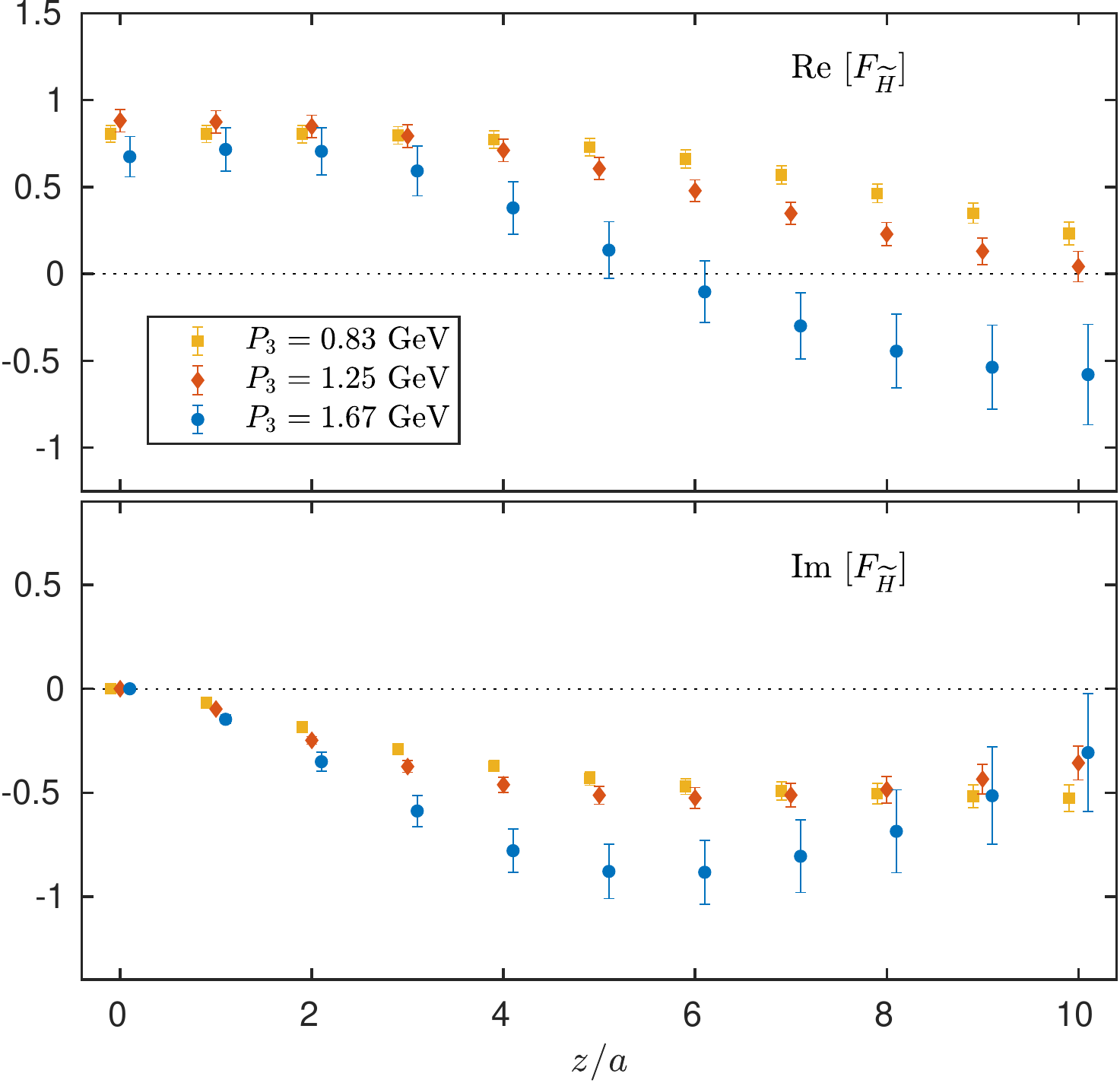}
\vspace*{-0.35cm}
\hspace*{0.35cm}
\begin{minipage}{15cm}
\caption{Left panels: The bare matrix elements $h_{\gamma_3\gamma^5}$ using parity projector ${\cal P}_3$ for $t=-0.69$ GeV$^2$ and $\xi=0$. Right panels: The renormalized $F_{\widetilde{H}}$. The top (bottom) plots correspond to the real (imaginary) part of the matrix elements. Momenta $P_3=0.83,\,1.25,\,1.67$ GeV are shown with orange squares, red diamonds, blue circles, respectively.}
\vspace*{0.5cm}
\label{fig:ME_pol}. 
\end{minipage}
\end{figure}

It is interesting to compare the matrix elements at fixed $P_3$ for different values of the skewness, and therefore, different values of $t$. In Fig.~\ref{fig:E_H_zeroxi_1over3}, we show $F_H$ and $F_E$ at $P_3=1.25$ GeV, with $\{t,\xi\}=\{$-0.69 GeV$^2,0\}$ and  $\{t,\xi\}=\{$-1.39 GeV$^2,1/3\}$. The real part of $F_E$ shows the largest sensitivity to such a simultaneous change of $t$ and $\xi$. In addition, the imaginary part of $F_H$ shows significant dependence on $\{t,\xi\}$ for $z/a>3$. 
The case of $\widetilde{F}_H$ is shown in the left panel of Fig.~\ref{fig:hel_xi_1over3}, where we observe large differences between $\{t,\xi\}=\{$-0.69 GeV$^2,0\}$ and $\{t,\xi\}=\{$-1.39 GeV$^2,1/3\}$. For completeness, we show $\widetilde{F}_E$ for $\xi=1/3$ in the right panel of Fig.~\ref{fig:hel_xi_1over3}.

\begin{figure}[h!]
\vspace*{0.5cm}
\includegraphics[scale=0.54]{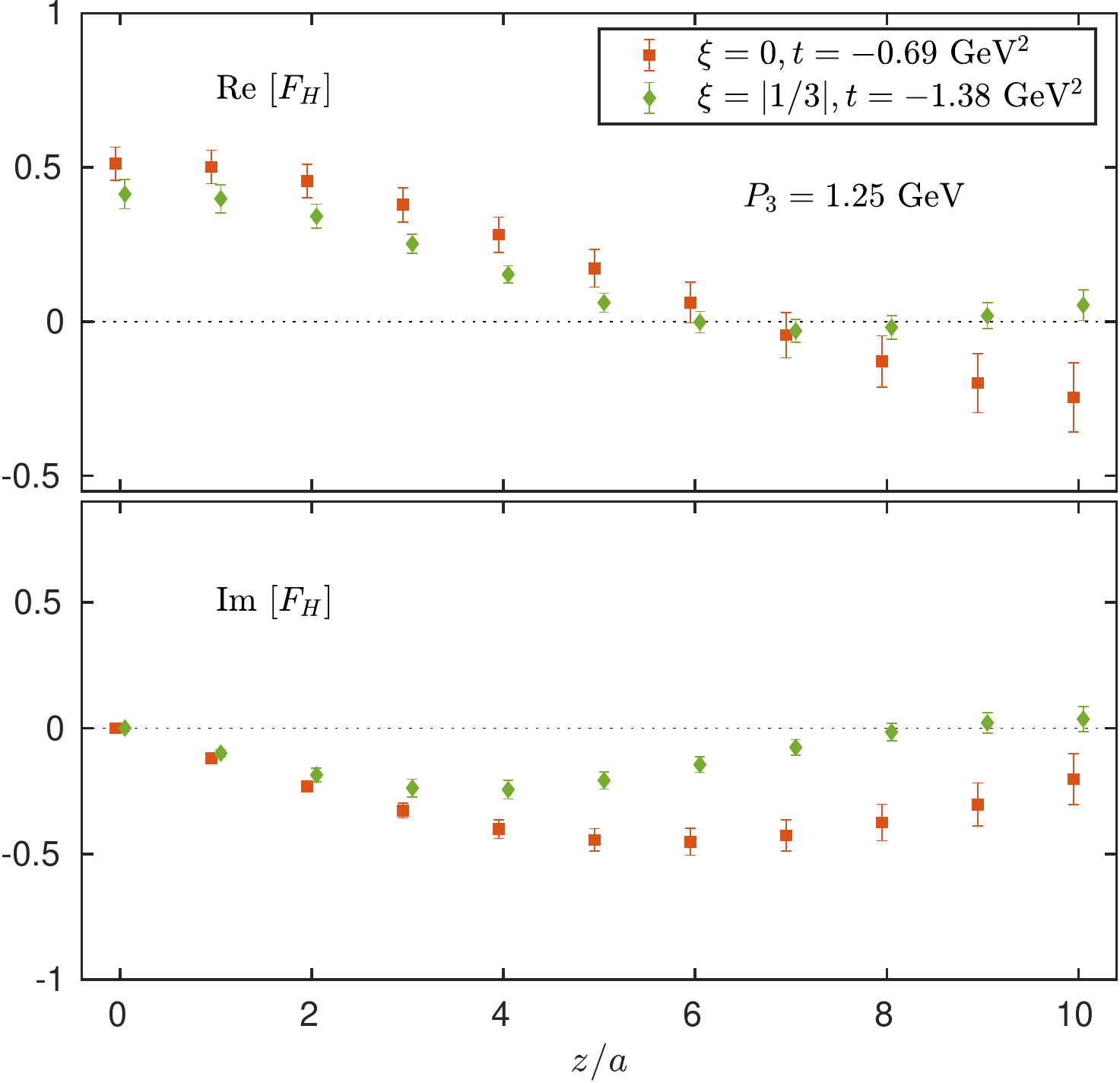}\hspace{0.2cm}
\includegraphics[scale=0.54]{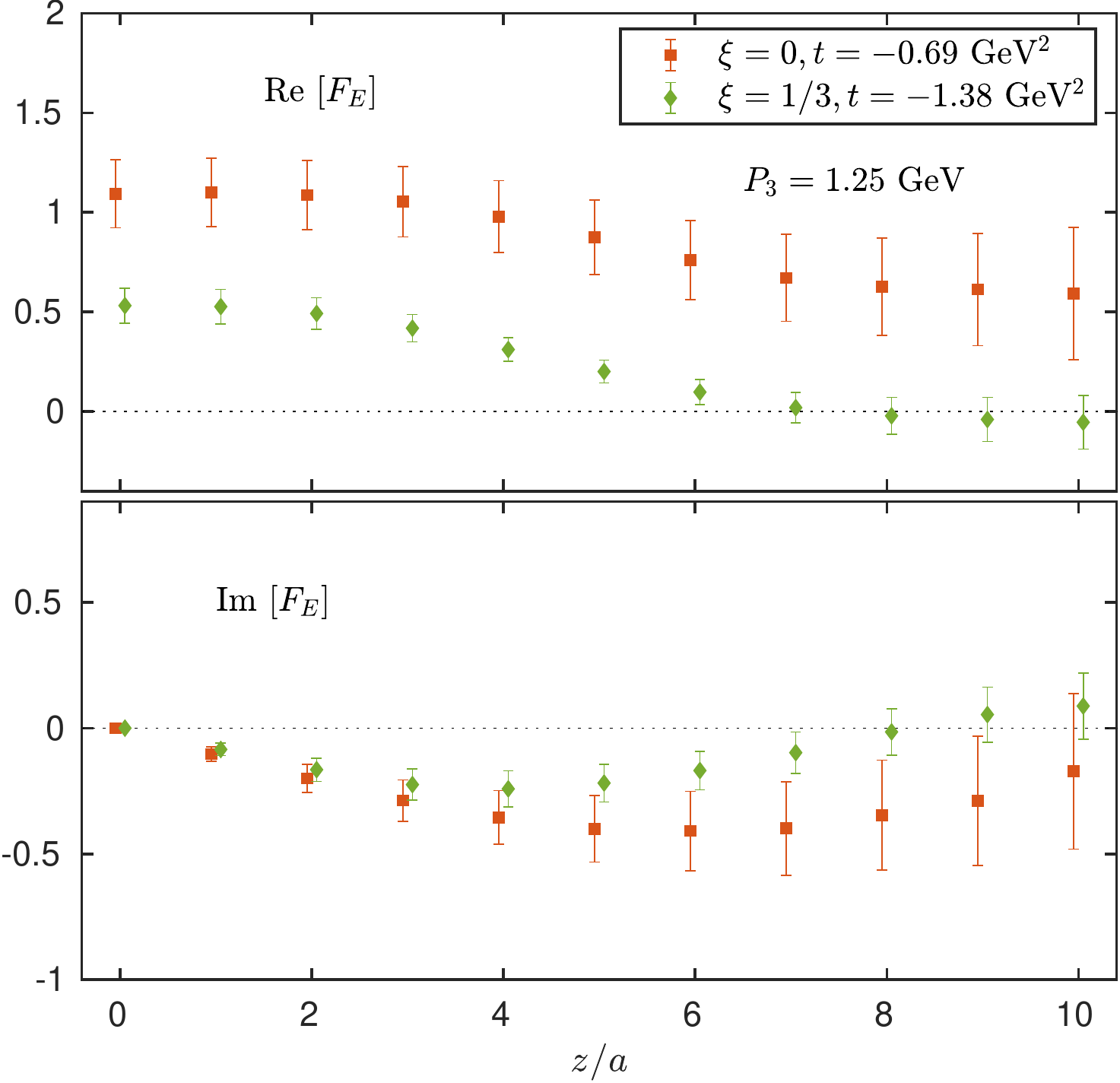}
\vspace*{-0.35cm}
\hspace*{0.35cm}
\begin{minipage}{15cm}
\caption{Renormalized quantities $F_H$ and $F_E$ for $P_3=1.25$ GeV, at $t=-0.69$ GeV$^2$, $\xi=0$ (red squares) and $t=-1.39$ GeV$^2$, $\xi=|1/3|$ (green diamonds).}
\label{fig:E_H_zeroxi_1over3}
\vspace*{0.5cm}
\end{minipage}
\end{figure}

\begin{figure}[h!]
\includegraphics[scale=0.54]{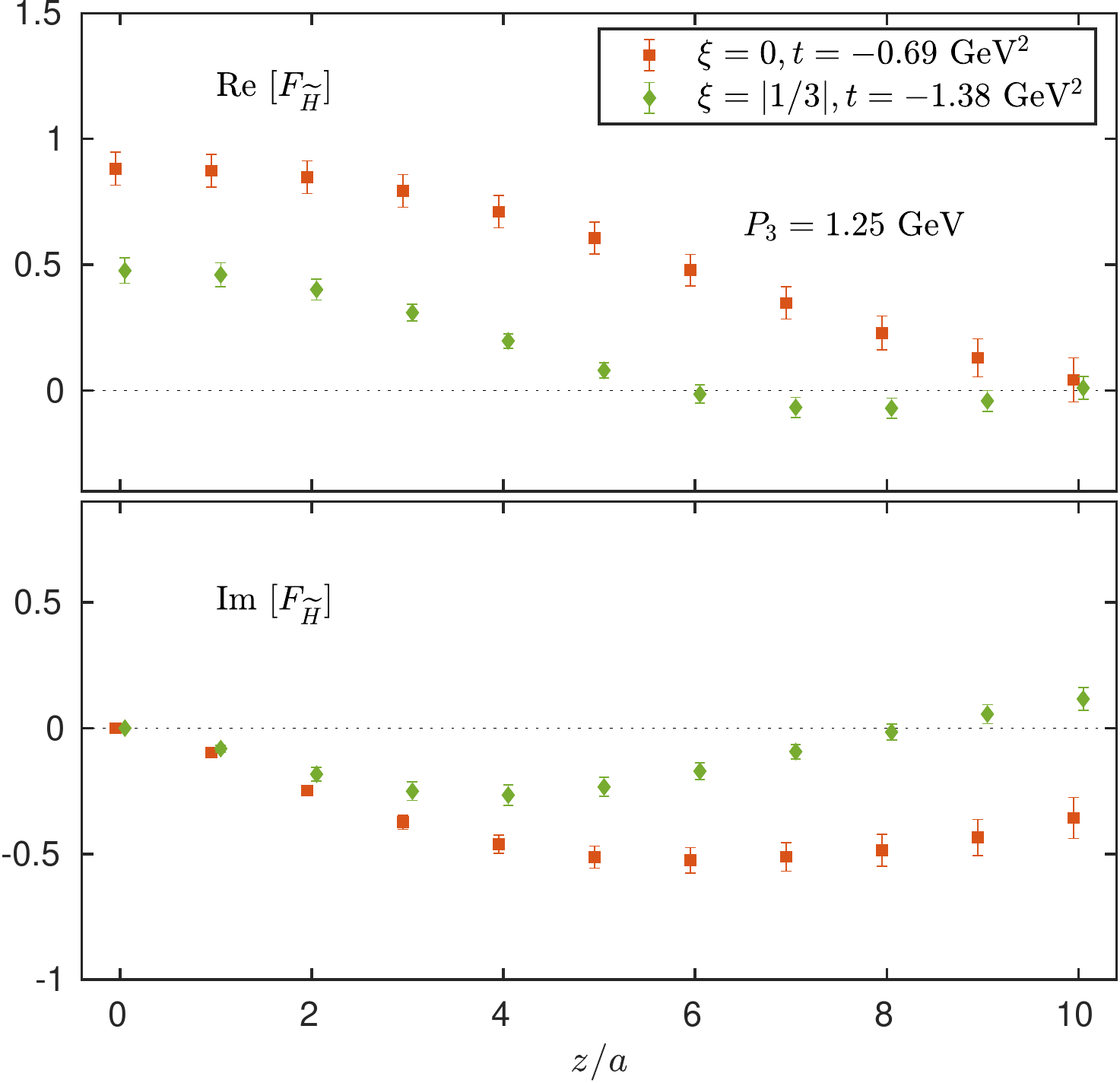}\hspace{0.2cm}
\includegraphics[scale=0.54]{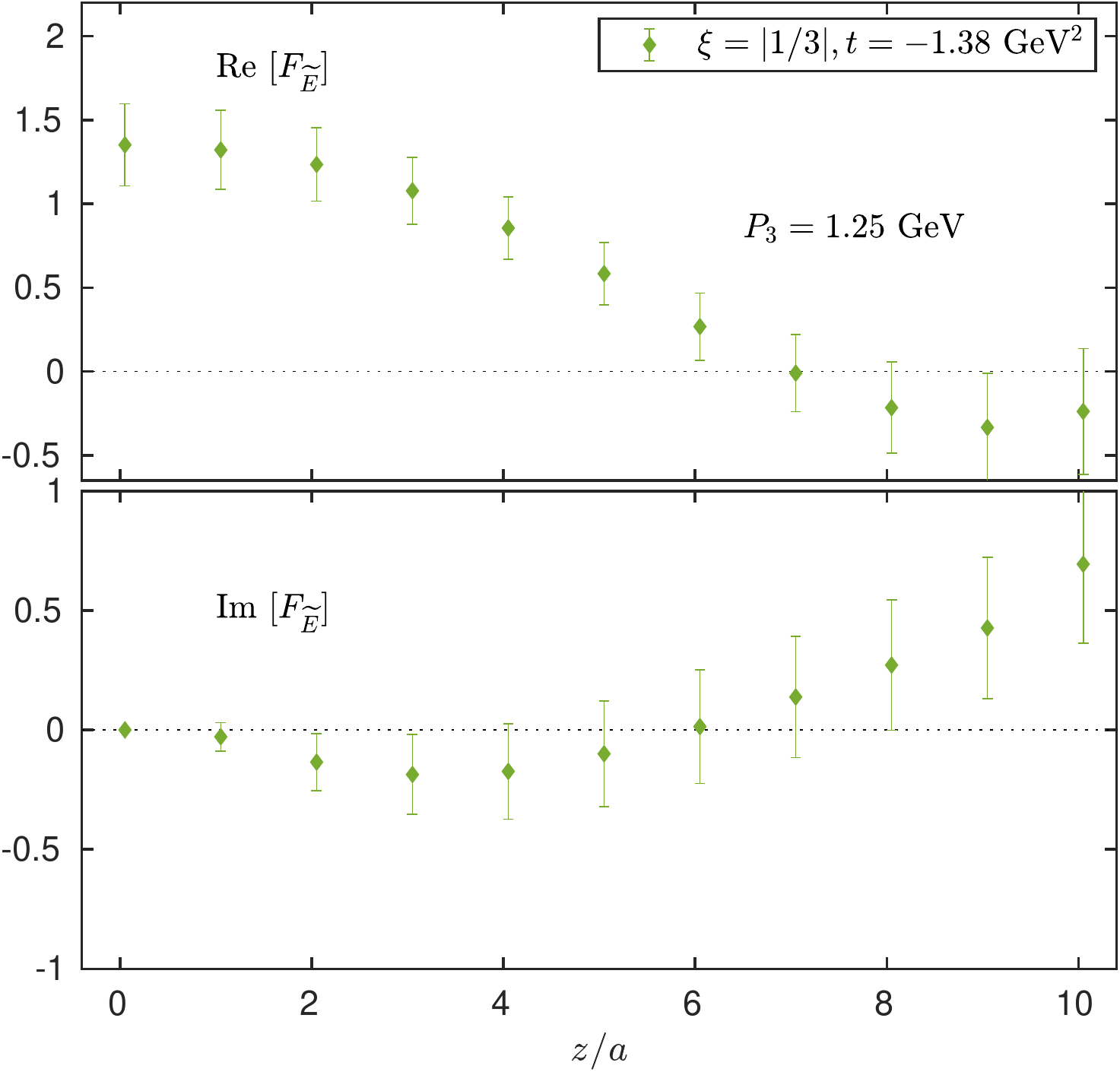}
\vspace*{-0.35cm}
\hspace*{0.35cm}
\begin{minipage}{15cm}
\caption{Renormalized $F_{\widetilde{H}}$ and $F_{\widetilde{E}}$ for $P_3=1.25$ GeV, at $t=-0.69$ GeV$^2$, $\xi=0$ (red squares) and $t=-1.39$ GeV$^2$, $\xi=|1/3|$ (green diamonds).}
\label{fig:hel_xi_1over3}
\vspace*{0.5cm}
\end{minipage}
\end{figure}

We now move on to the discussion of the final GPDs, in particular their convergence in momentum for zero skewness ($t=-0.69$ GeV$^2$). 
We compare the unpolarized GPDs for $P_3=0.83,\,1.25,\,1.67$ GeV in Fig.~\ref{fig:H_E_GPD_mom_depend}. The $H$-GPD has negligible $P_3$-dependence for every region of $x$, while the $E$-GPD exibits convergence between the two largest momenta for $x>0$, which is of main interest. We note that the statistical errors on $E$-GPD are larger than those of the $H$-GPD, a feature already observed in $F_E$ (see Fig.~\ref{fig:ME_unpol}). 
In Fig.~\ref{fig:Htilde_GPD_mom_depend}, we show the momentum dependence of the $\widetilde{H}$-GPD.
We observe that the relatively large differences between the renormalized $F_{\widetilde{H}}$ for the lowest two momenta and $P_3=1.67$ GeV are compensated by the matching procedure, indicating final convergence within the reported statistical uncertainties.
Thus, this conclusion holds for both the unpolarized and the helicity $H$-GPD.

\begin{figure}[h!]
\vspace*{.5cm}
\centering
\includegraphics[scale=0.54]{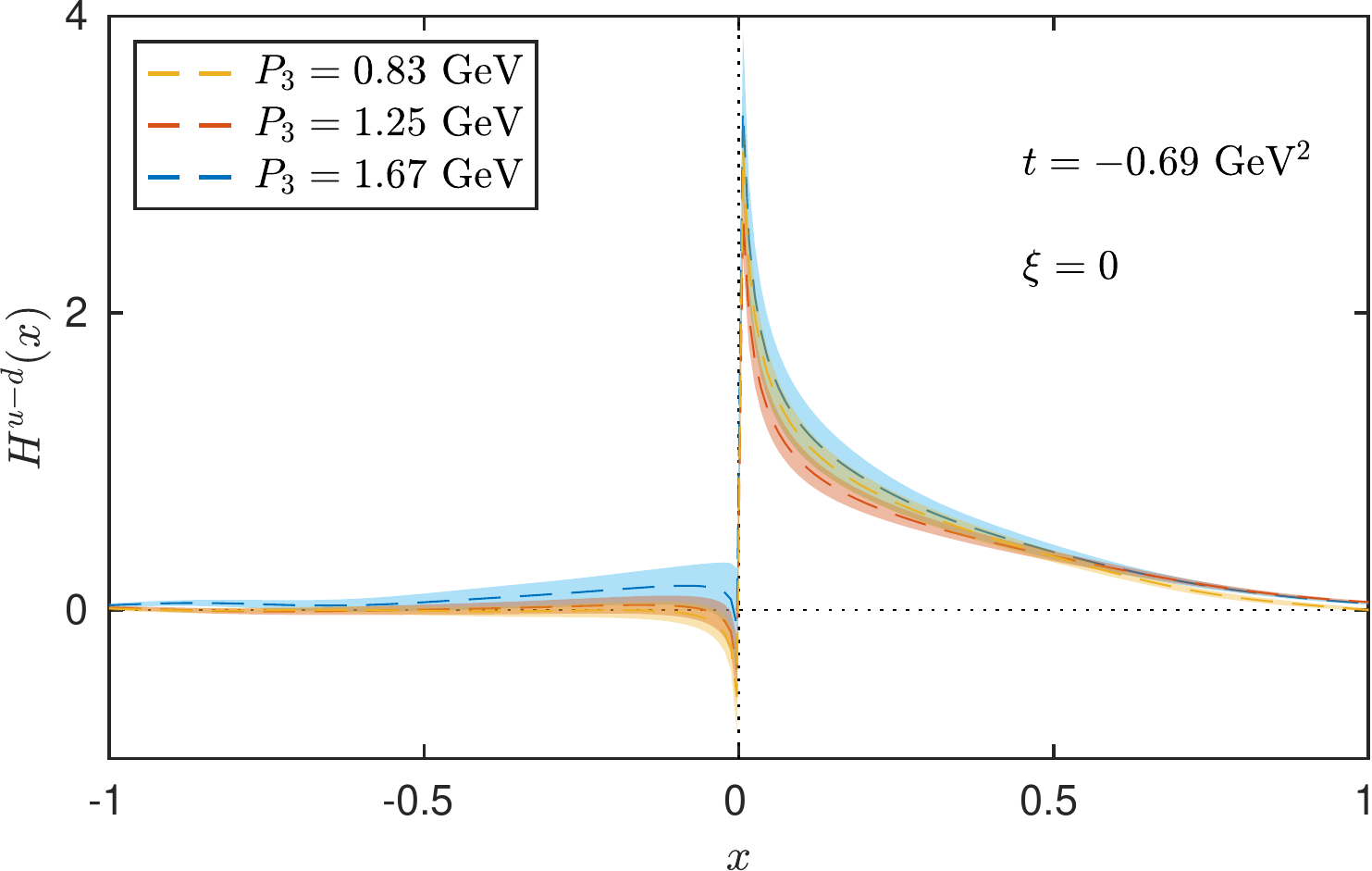}\hspace{0.2cm}
\includegraphics[scale=0.54]{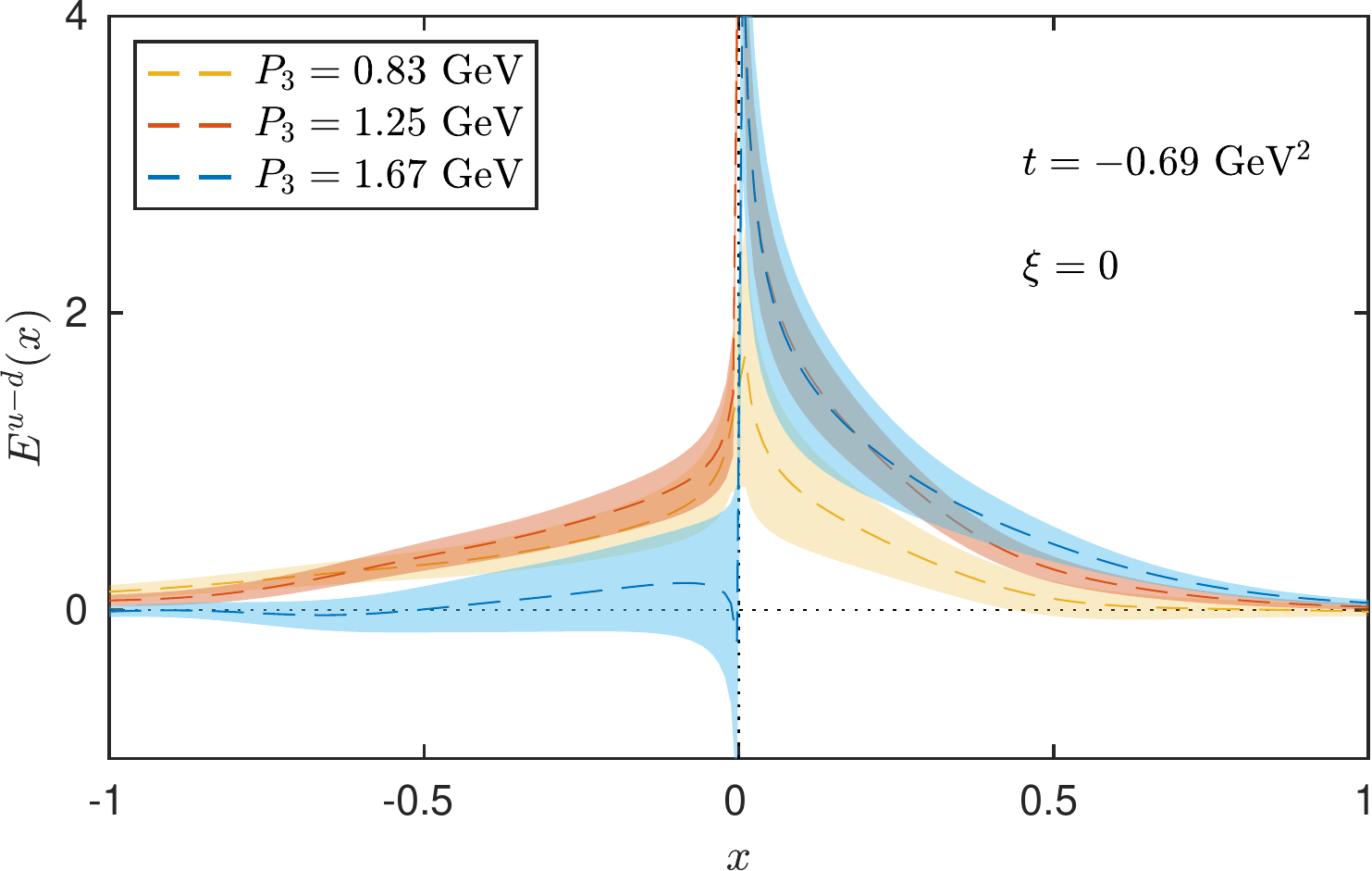}
\vspace*{-0.35cm}
\hspace*{0.35cm}
\begin{minipage}{15cm}
\caption{Momentum dependence of final $H$-GPD and $E$-GPD. Results on momenta $P_3=0.83,\,1.25,\,1.67$ GeV are shown with orange, red, blue band, respectively.}
\vspace*{0.5cm}
\label{fig:H_E_GPD_mom_depend}
\end{minipage}
\end{figure}

\begin{figure}[h!]
\vspace*{.5cm}
\centering
\includegraphics[scale=0.6]{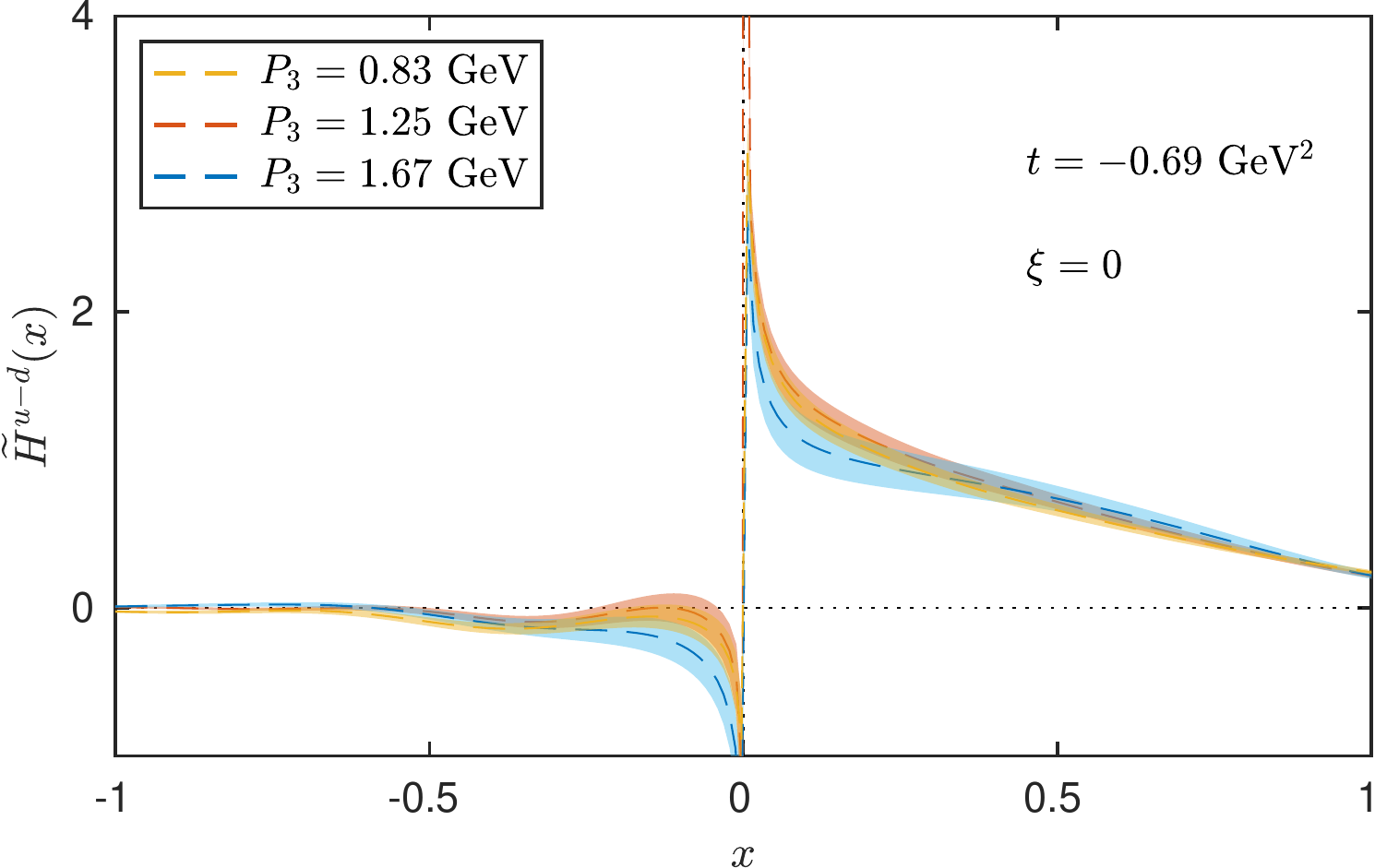}
\vspace*{-0.35cm}
\hspace*{0.35cm}
\begin{minipage}{15cm}
\caption{Momentum dependence of final $\widetilde{H}$-GPD using results for $P_3=0.83,\,1.25,\,1.67$ GeV, shown with orange, red, blue band, respectively.}
\vspace*{0.5cm}
\label{fig:Htilde_GPD_mom_depend}
\end{minipage}
\end{figure}

In Fig.~\ref{fig:H_E_comp_xi1over3} we provide a comparison between the $H$- and $E$-GPDs (left panel), and $\widetilde{H}$- and $\widetilde{E}$-GPDs (right panel).
In the unpolarized case, $H$- and $E$-GPDs are compatible with each other in the quark region.
However, the Pauli FF, corresponding to the $x$-integral of the $E$-GPD, is considerably larger than the Dirac FF (integral of $H$-GPD) at this momentum transfer and this is achieved by the larger values of the $E$-GPD in the antiquark region.
For the helicity case, the $\widetilde{E}$-GPD is significantly larger than the $\widetilde{H}$-GPD, which reflects the fact that the axial-vector FF $G_P$ is found to be a factor $\approx3$ larger than the $G_A$ at this momentum transfer, in a lattice setup with similar parameters \cite{Alexandrou:2013joa}.
We also note that the integrals of $H$-,$E$-,$\widetilde{H}$- and $\widetilde{E}$-GPDs extracted in this work are all compatible with their respective FFs obtained in Ref.~\cite{Alexandrou:2013joa}.

\begin{figure}[h!]
\vspace*{.5cm}
\centering
\includegraphics[scale=0.54]{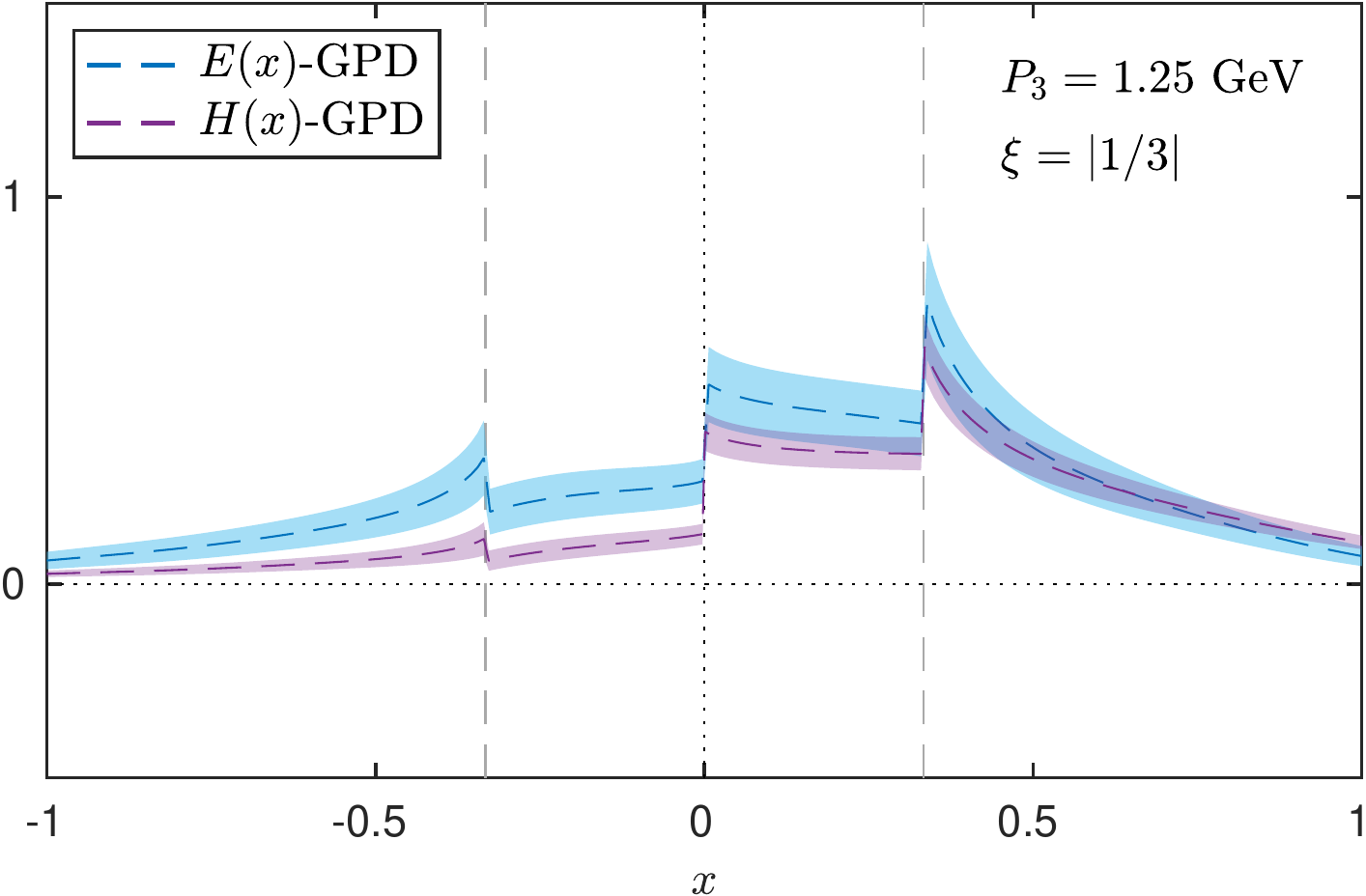}\hspace{0.2cm}
\includegraphics[scale=0.54]{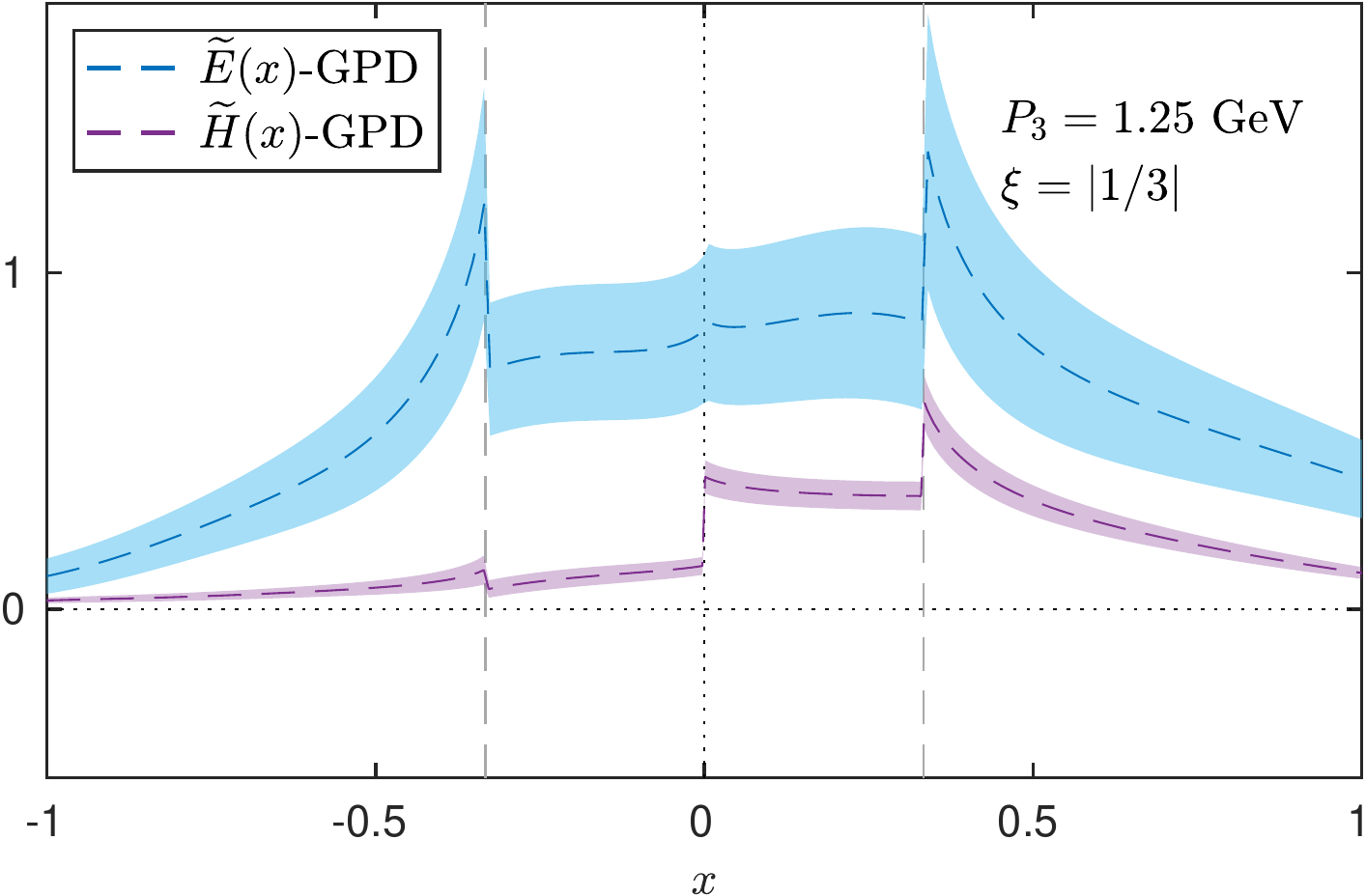}
\vspace*{-0.35cm}
\hspace*{0.35cm}
\begin{minipage}{15cm}
\caption{Left: Comparison of $H$-GPD (violet band) and $E$-GPD (blue band) for $P_3=,1.25$ GeV and $\xi=1/3$. Right: Same as left plot for helicity GPDs.}
\vspace*{0.5cm}
\label{fig:H_E_comp_xi1over3}
\end{minipage}
\end{figure}

\end{widetext}

\vspace*{5cm}
\bibliography{references}

\end{document}